\documentclass[12pt]{article}
\usepackage{jheppub_modified_green}

\usepackage{epsfig}
\usepackage{amsfonts}
\usepackage{amssymb,esint}
\usepackage{enumitem}
\usepackage{amsthm}
\usepackage{subfig}
\usepackage{bm}
\usepackage{hyperref}
\usepackage{mathrsfs}
\usepackage{bbm}
\usepackage{cancel}
\usepackage{xcolor}      
\usepackage{relsize}
\usepackage{float, slashed, graphicx, amssymb, amsmath}
\usepackage{shuffle}
\usepackage{pgfplots}
\usepackage{tikz-feynman}
\tikzfeynmanset{compat=1.1.0}
\tikzfeynmanset{graviton/.style={circle, draw=green!60, fill=green!5, very thick, minimum size=7mm}}
\tikzfeynmanset{codot/.style={/tikz/shape=circle,/tikz/fill=white,/tikz/minimum size=0.1cm,/tikz/inner sep=1.8pt}}
\tikzfeynmanset{myblob/.style={/tikz/shape=rectangle,/tikz/fill=red,/tikz/minimum size=0.2cm,/tikz/inner sep=1.8pt} }
\tikzfeynmanset{ghc/.style={/tikz/shape=circle,/tikz/fill=white,/tikz/minimum size=0.05cm,} }
\tikzfeynmanset{HV/.style={/tikz/shape=circle,/tikz/fill={rgb:black,1;white,2},/tikz/minimum size=0.1cm,} }
\tikzfeynmanset{GR/.style={/tikz/shape=ellipse,/tikz/fill={rgb:black,1;white,2},/tikz/minimum size=0.3cm,} }
\tikzfeynmanset{GR2/.style={/tikz/shape=ellipse,/tikz/fill={rgb:black,1;white,2},/tikz/minimum width = 1.2cm, 
    /tikz/minimum height = 3.4cm} }
\tikzset{box/.pic={\filldraw[fill=black]  (0,0) circle (2.5pt); \filldraw [fill=black] (0.5,0) circle (2.5pt); \draw [line width=5pt] (0,0) -- (0.5,0);}}
\usetikzlibrary{arrows.meta} % decent arrow in the graphs
\usetikzlibrary{calc}
\usetikzlibrary{decorations.pathmorphing}
\usetikzlibrary{decorations.pathreplacing}
\usetikzlibrary{decorations.markings}
\tikzset{
   vector2/.style={decorate, decoration={snake, amplitude=1pt, segment length=6pt}, draw,double},
   vector/.style={decorate, decoration={snake, amplitude=1pt, segment length=6pt}, draw},
	provector/.style={decorate, decoration={snake,amplitude=2.5pt}, draw},
	antivector/.style={decorate, decoration={snake,amplitude=-2.5pt}, draw},
    fermion/.style={draw=black, postaction={decorate},
        decoration={markings,mark=at position .55 with {\arrow[draw=black]{>}}}},
    fermionbar/.style={draw=black, postaction={decorate},
        decoration={markings,mark=at position .55 with {\arrow[draw=black]{<}}}},
    fermionnoarrow/.style={draw=black},
    gluon/.style={decorate, draw=black,
        decoration={coil,amplitude=4pt, segment length=5pt}},
    scalar/.style={dashed,draw=black, postaction={decorate},
        decoration={markings,mark=at position .55 with {\arrow[draw=black]{>}}}},
    scalarbar/.style={dashed,draw=black, postaction={decorate},
        decoration={markings,mark=at position .55 with {\arrow[draw=black]{<}}}},
    scalarnoarrow/.style={dashed,draw=black},
    electron/.style={draw=black, postaction={decorate},
        decoration={markings,mark=at position .55 with {\arrow[draw=black]{>}}}},
	bigvector/.style={decorate, decoration={snake,amplitude=4pt}, draw},
}
\tikzset{cross/.style={cross out, draw, 
         minimum size=2*(#1-\pgflinewidth), 
         inner sep=0pt, outer sep=0pt}}

\tikzstyle{block} = [draw, rectangle, 
    minimum height=3em, minimum width=6em]

\usepackage[T1]{fontenc} % if needed

\def\sc#1{\overline{#1}}
\makeatletter
\newcommand \UPlus {\mathop {\operator@font \uplus }\limits }
\makeatother
\makeatletter
\newcommand \Bigcup {\mathop {\operator@font \bigcup }\limits }
\makeatother
  \def\LabelNote#1{}%\smash{\hbox to\phipt{\raise1ex\hbox{\tiny[#1]}\hss}}}
 \def\Label#1{\label{#1}%
  \smash{\hbox to\phipt{\raise1ex\hbox{\tiny[#1]}\hss}}}
  
  \def\mdot{{\cdot}}

\newcommand{\cM}{\mathcal{M}}

\newcommand{\pb}{\bar p}

\def\nn{\nonumber}

\def\spa#1.#2{\left\langle#1\,#2\right\rangle}
\def\spb#1.#2{\left[#1\,#2\right]}

\def\be{\begin{equation}}
\def\ee{\end{equation}}
\def\bea{\begin{eqnarray}}
\def\eea{\end{eqnarray}}  
\allowdisplaybreaks

\newcommand{\npre}{\mathcal{N}}

\newcommand{\tF}{\widetilde F}

\usepackage{amssymb,amsmath}
\usepackage{mathtools} %added for coloneqq
\usepackage{cancel} %crossing-out terms
\usepackage{graphicx} %Lets us rotate symbols

\usepackage{hyperref}
\hypersetup{
	colorlinks=true,
	linktoc=page,
	citecolor=cadmiumgreen,
	linkcolor=cadmiumgreen,
	urlcolor=blue} 
\definecolor{americanrose}{rgb}{1.0, 0.01, 0.24}
\definecolor{cadmiumgreen}{rgb}{0.0, 0.42, 0.24}

\usepackage[colorinlistoftodos]{todonotes}
\newcommand{\Cdot}{{\cdot}} % dot with less space
\def\nn{\nonumber}

\begin{document}

%%% --- Title page --- %%%
\title{Dynamics of Spinning Binary at 2PM}
\author{Gang Chen$\mbox{}^{a}$,}
\author{Tianheng Wang$\mbox{}^{b,c}$}
\affiliation{$\mbox{}^{a}$Niels Bohr International Academy,
Niels Bohr Institute, University of Copenhagen,\\
Blegdamsvej 17, DK-2100 Copenhagen \O, Denmark}
\affiliation{$\mbox{}^{b}$Center for Theoretical Physics, Seoul National University, \\
1 Gwana-ro, Gwanak-gu, 08826, Seoul, South Korea}
\affiliation{$\mbox{}^{c}$Institute of Theoretical Physics, Chinese Academy of Sciences, \\
55 Zhongguancun Road East, Haidian District, 100190,  Beijing, China}
\emailAdd{gang.chen@nbi.ku.dk}
\emailAdd{tianhengwang@snu.ac.kr}

\begin{flushright}
	SNUTP24-002
\end{flushright}

\abstract{
We consider the covariant proposal for the gravitational Compton amplitude for a Kerr black hole. Employing the covariant three- and four-point Compton amplitudes, we assemble the classical one-loop integrand on the maximal cut at all orders in spin, utilizing the method of unitarity. Expanding in powers of spin, we evaluate the one-loop amplitude up to $\mathcal O(G^2 a^8)$. Supplemented with extra contact contributions derived from the far-zone data of the Teukolsky solutions, the one-loop amplitude is in agreement with results available in the literature. We showcase the classical eikonal in the aligned-spin case at $\mathcal O(G^2 a^7)$.
}

\vspace{-2.6cm}

%\pagenumbering{roman}
\maketitle

\flushbottom
 \tableofcontents
%\newpage 
%please don't remove this newpage otherwise the intro has no page number

\section{Introduction}
The successful detection of gravitational waves~\cite{LIGOScientific:2016dsl,LIGOScientific:2016aoc,LIGOScientific:2016sjg,LIGOScientific:2017bnn,LIGOScientific:2017vwq} has inspired an explosion of developments in the studies of black hole mergers. Besides numerical relativity~\cite{Pretorius:2005gq,Campanelli:2005dd,Baker:2005vv} and the formal approaches focusing on the post-Newtonian expansion~\cite{Buonanno:1998gg,Goldberger:2004jt,Kol:2007rx,Gilmore:2008gq,Foffa:2011ub,Foffa:2016rgu,Porto:2017dgs,Blumlein:2019zku,Foffa:2019rdf,Foffa:2019yfl,Blumlein:2020pog,Blumlein:2020znm,Bini:2020nsb,Bini:2020hmy,Blumlein:2021txe,Blumlein:2020pyo,Foffa:2020nqe,Blumlein:2021txj,Kim:2021rfj,Cho:2022syn}, various amplitude-based approaches~\cite{Bjerrum-Bohr:2018xdl,Cheung:2018wkq,Bern:2019nnu,Bern:2019crd,Neill:2013wsa,Cristofoli:2021vyo,Bern:2020buy,Bern:2021dqo,Bjerrum-Bohr:2002gqz} have proven highly effective in computing higher-order corrections in the Post-Minkowskian (PM) expansion and revealing the underlying structures of gravitational interactions.

These amplitude-based approaches view the dynamics of heavy bodies interacting via weak gravity as that of scattering processes in the classical regime where the gravitons are soft. A multitude of methods~\cite{Kosower:2018adc,Maybee:2019jus,Brandhuber:2021kpo,Brandhuber:2021eyq,Mogull:2020sak,Jakobsen:2021smu,Jakobsen:2022fcj,Wang:2022ntx,Parra-Martinez:2020dzs,DiVecchia:2021bdo,Heissenberg:2021tzo,DiVecchia:2022nna,Damgaard:2021ipf,Kol:2021jjc,Bjerrum-Bohr:2021wwt,Bern:2021xze,Lee:2023vdy,Lee:2023zuu} have been devised to single out the classical order from the full (quantum) amplitudes. Techniques for computing loop amplitudes efficiently facilitate computations at high PM orders. At the point of writing, black hole scattering observables have been computed up to 4PM~\cite{Dlapa:2021npj,    Dlapa:2022lmu,Dlapa:2023hsl,Jakobsen:2023pvx,Bern:2021yeh,Bern:2021dqo,Damgaard:2023ttc} for both non-spinning and spinning binaries and partial results are obtained at 5PM in the non-spinning case~\cite{Driesse:2024xad}.

Incorporating spin degrees of freedom is non-trivial, both conceptually and computationally. Challenges appear even at tree level. Although the three-point amplitudes of two massive particles and one graviton are classified in~\cite{Arkani-Hamed:2017jhn}, complications enter at four points, namely the gravitational Compton scattering.
On the one hand, fundamental questions in quantum field theories with higher-spin particles need to be addressed~\cite{Zinoviev:2001dt,Zinoviev:2006im,Zinoviev:2009hu,Zinoviev:2008ck,Ochirov:2022nqz,Cangemi:2022abk}, towards a full understanding of these amplitudes. On the other hand, the concept of double copy~\cite{Bern:2008qj,Bern:2010ue,Bern:2019prr} and general considerations on symmetry and locality requirements allow us to bootstrap the Compton amplitudes of arbitrary spins~\cite{Bjerrum-Bohr:2023jau,Bjerrum-Bohr:2023iey}, despite incomplete knowledge of the underlying theory. These results can then be tested against approaches developed from other perspectives such as the Teukolsky equation~\cite{Teukolsky:1973ha,Press:1973zz,Teukolsky:1974yv,Chia:2020yla,Bautista:2021wfy,Ivanov:2022qqt,Bautista:2022wjf,Saketh:2023bul,Bautista:2023sdf}.

Based on somewhat different assumptions, the Compton amplitude for a Kerr black hole has been computed both as an expansion in powers of spin \cite{Chung:2018kqs,Chung:2019duq,Chung:2020rrz,Chen:2021kxt,Aoude:2022trd,Aoude:2022thd,Bern:2022kto,Bern:2023ity,Menezes:2022tcs,FebresCordero:2022jts,Cangemi:2022bew,Haddad:2023ylx,DeAngelis:2023lvf,Chen:2023qzo} and as a resummed function~\cite{Cangemi:2023ysz,Cangemi:2023bpe,Bjerrum-Bohr:2023jau,Bjerrum-Bohr:2023iey}. They all agree up to the quartic order in the spin expansion, where the spin-shift symmetry is expected to hold. Going to higher orders in spin, they deviate from one another, as the nuances in their respective formalism and assumptions become important. These results provide a pool of data for future investigations to analyze and clarify, which in turn will help to reveal the structures of the underlying theory. 
The Compton amplitude for a Kerr black hole is also an important ingredient in the calculation of loop amplitudes, which describe the dynamics of Kerr binaries at higher PM orders~\cite{Vines:2017hyw,Guevara:2018wpp,Chung:2020rrz,Kosmopoulos:2021zoq,Scheopner:2023rzp,Alessio:2022kwv,Elkhidir:2023dco,Bini:2023fiz,Chen:2021kxt,Brandhuber:2023hhl,Luna:2023uwd,Lee:2023nkx,Buonanno:2024vkx}. 

A complementary line of research~\cite{Jakobsen:2021lvp,Jakobsen:2021zvh,Jakobsen:2022zsx,Jakobsen:2023ndj,Jakobsen:2023hig,Shi:2021qsb,Comberiati:2022cpm,Gonzo:2023goe} based on the worldline quantum field theory (WQFT) formalism~\cite{Mogull:2020sak} provides a fast-track to high PM calculations for physical observables and generating functions, up to quadratic orders in spin. Recent development in the worldline formalism focuses on the radial action, which produces scattering observables at 6PM beyond the quadratic order in certain kinematic regions~\cite{Gonzo:2024zxo}.
The gravitational waveform in the spinning case is discussed in ~\cite{DeAngelis:2023lvf,Brandhuber:2023hhl,Aoude:2023dui,Bohnenblust:2023qmy} up to the quartic order in spin. Besides Einstein gravity, related works on the effects of spin in gauge~\cite{Skvortsov:2023jbn,Kim:2024grz} and other gravitational backgrounds~\cite{Brandhuber:2024bnz,Bern:2024adl} have been carried out. 
Beyond conservative dynamics, tidal and radiation effects have been explored from both amplitude approaches and worldline-based formalisms~\cite{Kalin:2020fhe,Kalin:2020lmz,Riva:2021vnj,Dlapa:2021vgp,Goldberger:2020fot,Kalin:2020mvi,Bhattacharyya:2024aeq}.
Recent developments on the self-force expansion~\cite{Kosmopoulos:2023bwc,Cheung:2023lnj,Wilson-Gerow:2023syq,Jakobsen:2023tvm,Klemm:2024wtd} may also shed light on the dynamics in curved backgrounds.

In this paper, we consider one particular proposal for the Compton amplitude~\cite{Bjerrum-Bohr:2023jau,Bjerrum-Bohr:2023iey}, which is obtained using the double-copy and bootstrap. In section~\ref{sec:ClassicalCompton}, we review its structures and make comparisons with other proposals at the level of their spin expansions. 
In section~\ref{sec:2PMAngle}, we compute the one-loop amplitude using this Compton amplitude from the leading singularity and the resulting 2PM eikonal phase.

\section{Classical Compton amplitudes}\label{sec:ClassicalCompton}
We begin this section with a review of the classical three-point~\cite{Arkani-Hamed:2019ymq,Bern:2020buy,Johansson:2019dnu,Bjerrum-Bohr:2023jau} and four-point Compton amplitudes~\cite{Bjerrum-Bohr:2023jau,Bjerrum-Bohr:2023iey} of two massive particles with spins minimally coupled to gravity at tree level. The Compton amplitude follows from the double-copy and factorization requirements and includes contact terms. The contact terms match with physical data at low orders in spin and display certain empirical properties, which allows for an extrapolation. This Compton amplitude is conjectured to hold at all orders in spin. In the second part of this section, we discuss the comparison with the Compton amplitudes computed from other approaches~\cite{Cangemi:2023bpe,Bautista:2023sdf}. The differences in the contact terms can be interpreted as being related to the internal structures of the Kerr black hole. Moreover, we demonstrate a simple procedure to obtain expressions in covariant variables for these extra contact terms. 

\subsection{Covariant Compton amplitude to arbitrary spin order}\label{sec:Compton}
Here we review the structures of the classical amplitudes of a heavy particle emitting gravitons.
Throughout this paper, we restrict our discussions to the kinematics in the heavy-mass limit~\cite{Georgi:1990um, Luke:1992cs, Neubert:1993mb, Manohar:2000dt,Damgaard:2019lfh,Brandhuber:2021eyq}. The incoming and outgoing massive momenta typically are parameterised as $\bar p = m v$ and $\bar p' = (m v - q)$ where $q^\mu$ denotes the total momenta of the emitted graviton(s). The on-shell conditions $\bar p^2 = \bar p^{\prime2} =(mv - q)^2= m^2$ yield $v^2 =1 $ and $v\mdot q=0$.

The three-point amplitude of two massive particles of arbitrary masses and spins and a graviton is first given in \cite{Conde:2016izb,Conde:2016vxs,Arkani-Hamed:2019ymq} by considering general symmetry constraints. Restricting to the minimal coupling and taking the heavy-mass limit, the amplitude corresponds to the Kerr black hole of an arbitrary classical spin.  
A compact, covariant form of this three-point amplitude is found in \cite{Bern:2020buy,Johansson:2019dnu,Bjerrum-Bohr:2023jau}    
\begin{equation}\label{threepointgrav}
	\begin{tikzpicture}[baseline={([yshift=-0.8ex]current bounding box.center)}]\tikzstyle{every node}=[font=\small]	
\begin{feynman}
    	 \vertex (a) {\( \sc p\)};
    	 \vertex [right=1.5cm of a] (f2)[dot]{};
    	 \vertex [right=1.5cm of f2] (c){$\sc p'$};
    	 \vertex [above=1.3cm of f2] (gm){$\varepsilon_1, p_1$};
    	  \diagram* {
(a) -- [thick] (f2) --  [thick] (c),
    	  (gm)--[photon,ultra thick](f2)
    	  };
    \end{feynman}  
    \end{tikzpicture}\,\,\,\,\,
      \cM_3(1,\sc p',\sc p)= -i\kappa (\pb\Cdot\varepsilon_1)({\mathsf w}\mdot\varepsilon_1)\,,
\end{equation}
where  ${\mathsf w}^\mu \coloneqq\cosh(x_1)\pb^\mu-iG_1(x_1)(p_1\Cdot S)^\mu$, with $G_1(x_1)=\frac{\sinh(x_1)}{x_1}$, $x_1=a \mdot p_1$ and $(p_1 \mdot S)^\mu = p_{1\nu}S^{\nu\mu}$. Here $S^{\mu\nu}$ denotes the spin tensor which describes the classical spin of the black hole and we have introduced the spin-length vector
$a^\mu$, with $S^{\mu\nu} = -\epsilon^{\mu\nu\rho\sigma} \bar{p}_\rho a_\sigma$. 
As $G_1(x_1)$ is an entire function, (\ref{threepointgrav}) is local in the sense that it is free of unphysical poles involving $x_1$ in the denominator once expanded in powers of the spin vector.

At four points, we have the classical gravitational Compton scattering of two gravitons and the Kerr black hole, depicted below.  
\begin{align}
	\begin{tikzpicture}[baseline={([yshift=-0.8ex]current bounding box.center)}]\tikzstyle{every node}=[font=\small]	
\begin{feynman}
    	 \vertex (a) {\( \sc p\)};
    	 \vertex [right=1.9cm of a] (f2)[GR]{$~~\mathbf{S}~~$};
    	 \vertex [right=1.9cm of f2] (c){$\sc p'$};
    	 \vertex [above=1.3cm of f2] (gm){};
    	 \vertex [left=0.8cm of gm] (g2){$\varepsilon_1,p_1$};
    	  \vertex [right=0.8cm of gm] (g20){$\varepsilon_2,p_2$};
    	  \diagram* {
(a) -- [thick] (f2) --  [thick] (c),
    	  (g2)--[photon,ultra thick](f2),(g20)--[photon,ultra thick](f2)
    	  };
    \end{feynman}  
    \end{tikzpicture}\, .
\end{align}
A covariant form of the Compton amplitude for a Kerr black hole is obtained in \cite{Bjerrum-Bohr:2023iey,Bjerrum-Bohr:2023jau}. 
Schematically, this amplitude receives three types of contributions as follows, 
\begin{align}\label{amplitude structure}
    \cM_4(1,2, \sc p', \sc p)&=-{\npre_a(1,2, \sc p', \sc p)\,\npre_0(1,2, \sc p, \sc p')\over 2(p_1\mdot p_2)}+{\npre_{\rm r}(1,2, \sc p', \sc p)\over 4(\sc p\mdot p_1) (\sc p\mdot p_2)} +\npre_{\rm c}(1,2, \sc p', \sc p).
\end{align}
Here the first term, as its form suggests, is assembled from the double-copy procedure. The second term is designated to take care of the spin-flip effects. The last denotes the contact contributions which can not be determined from factorizations.
The Compton amplitude here is expressed in terms of the same set of variables as the three-point amplitude~\eqref{threepointgrav}, where the spin tensor satisfies $\bar{p}_\mu S^{\mu\nu}=0$.

The double-copy term, constructed in~\cite{Bjerrum-Bohr:2023jau}, is given by $\mathcal{N}_a$ denoting the kinematic numerator of the Yang-Mills Compton amplitude in the heavy-mass limit and its spinless counterpart $\mathcal N_0$. The numerator $\mathcal N_a$ takes the following form,
\begin{align}\label{eq:NaDC}
	&\npre_a(1,2, \sc 3, \sc 4)=-\frac{{\mathsf w}_1\mdot F_1\mdot F_2\mdot {\mathsf w}_2}{ (p_1\mdot \bar p)}+\frac{p_1\mdot \bar p-p_2\mdot \bar p}{2(p_1\mdot \bar p)}\times\\\
    & \Big(i G_2\left(x_1,x_2\right) (a\mdot F_1\mdot F_2\mdot S\mdot p_2)+i G_2(x_1,x_2) (a\mdot F_2\mdot F_1 \mdot S\mdot p_1)+i G_1\left(x_{12}\right) \text{tr}\left(F_1\mdot S\mdot F_2\right) \nn\\
      &+ G_1(x_1) G_1(x_2) \big( (a\mdot F_1\mdot \bar p) (a\mdot F_2\mdot p_1)\!-\!(a\mdot F_1\mdot p_2) (a\mdot F_2\mdot \bar p) -\!\frac{p_2\mdot \bar p\!-\!p_1\mdot \bar p}{2}  (a\mdot F_1\mdot F_2\mdot a)\big)\Big)\, ,\nn
\end{align}
where $x_i=a_i\mdot p$,~$i=1,2$ and $x_{12}=x_1+x_2$ and 
\begin{align}
	G_2(x_{1}, x_{2})&\equiv {1\over x_{2}}\Big({\sinh(x_{12})\over x_{12}}-\cosh(x_{2}) {\sinh(x_1)\over x_1}\Big).
\end{align}
$G_2(x_1,x_2)$ is also an entire function and hence renders $\npre_a$ free of unphysical poles in $x_1$, $x_2$ or $x_{12}$.
The scalar numerator $\mathcal{N}_0$ follows from the kinematic Hopf algebra \cite{Brandhuber:2021bsf,Brandhuber:2022enp,Chen:2022nei,Chen:2023ekh,Chen:2024gkj,Bjerrum-Bohr:2024fbt}
\begin{align}
	\npre_0(1,2, \sc p, \sc p')=-\Big(\!\frac{\bar p'\mdot F_1\mdot F_2\mdot \bar p'}{p_1\mdot \bar p'}\Big).
\end{align}
We have  $\mathcal{N}_0=\mathcal{N}_a|_{a\rightarrow 0}$. These numerators manifestly satisfy the crossing symmetries described by group algebra in actions~\cite{Chen:2019ywi,Chen:2021chy} as demanded by the color-kinematic duality. The double-copy term is consistent with all possible massless factorizations.

The double-copy construction dictates that the spin direction and the magnitude remain unchanged during the propagation. Hence, additional contributions need to be incorporated to produce the correct spin-flip effects, starting from the cubic order in spin. This is given by the second term in~\eqref{amplitude structure}. More detailed discussions on the spin-flip effects are given in~\cite{Bjerrum-Bohr:2023iey}. 
The spin-flip contribution is determined by the factorization requirement on the physical massive poles. The numerator $\mathcal N_r$ reads 
\begin{align}
	 \npre_{\rm r}(1,2, \sc p', \sc p)=&\!{\Big((\partial_{x_1}\!-\!\partial_{x_2})G_1(x_1) G_1(x_2)\Big)\over 4(\bar p\mdot p_1) (\bar p\mdot p_2)} \Big(\bar p\mdot p_2(\bar p^2 (a\mdot F_1\mdot F_2\mdot a) (a\mdot F_2\mdot F_1\mdot \bar p) \nn\\
      &+a^2 (\bar p\mdot F_1\mdot F_2\mdot \bar p) (a\mdot F_1\mdot F_2\mdot \bar p))\!-\! (1\leftrightarrow 2)\!\Big)\nn\\
     &+\Big({i(\partial_{x_1}\!-\!\partial_{x_2})G_2(x_1,x_2)\over 4(\bar p\mdot p_1) (\bar p\mdot p_2)}\Big) \Big((\bar p\mdot p_2) (a\mdot F_2\mdot F_1\mdot \bar p) ((a\mdot F_2\mdot \bar p) (a\mdot \tF_1\mdot \bar p)\nn\\
      &-(a\mdot F_1\mdot \bar p) (a\mdot \tF_2\mdot \bar p))+(1\leftrightarrow 2)\Big).
\end{align} 
The consistency conditions resulted from all the physical factorizations \cite{Chen:2021kxt} are satisfied. 

It is noted in~\cite{Bjerrum-Bohr:2023iey,Bjerrum-Bohr:2023jau} that the three-point amplitude, the double-copy and the spin-flip contributions to the Compton amplitude share the property that their \emph{degrees} are zero. Here the degree is defined as the maximal power of $1/\chi$, when $\chi$ parameterizes the scaling of the spin variable, namely $a^\mu\rightarrow \chi a^\mu$ and $\chi\rightarrow i\infty$. 

The contact contribution is not accessible via factorizations and instead is obtained by matching with physical data~\cite{Bautista:2022wjf,Bautista:2023szu} at $\mathcal O(a^4)$ and $\mathcal O(a^5)$. Such expressions also exhibit the characteristic above that their degrees are zero at low orders in spin. Imposing this as a general constraint, the following contact contribution is extrapolated to all orders in spin, 
\begin{align}
 \npre_{\rm c}(1,2, \sc p', \sc p)&=\!\Big(\!{(\partial_{x_1}\!-\!\partial_{x_2})^2\over 2!}{G_{1}(x_1)}{G_{1}(x_2)}\Big)\Big((a\mdot F_1\mdot \bar p) (a\mdot F_2\mdot \bar p) (a\mdot  F_1\mdot F_2\mdot a)\!\nn\\
      &-\!{a^2\over 2} ((a\mdot F_1 \mdot F_2\mdot p) (a\mdot F_2 \mdot F_1\mdot \bar p) + (a\mdot F_1 \mdot F_2\mdot a) (\bar p\mdot F_1 \mdot F_2\mdot \bar p))\!\Big)\\
     &\!\!\!+\Big({i(\partial_{x_1}-\partial_{x_2})^2\over 2!}{G_{2}(x_1,x_2)}\Big)\Big(\! -{1\over 2} \big((a\mdot F_1\mdot F_2\mdot a) (a\mdot F_2\mdot \bar p) (a\mdot \tF_1\mdot \bar p)\!-\!(1\leftrightarrow 2)\big)\Big).\nn
\end{align}

Putting together all three types of contributions, the covariant expression for the Compton amplitude~\eqref{amplitude structure} is manifestly gauge invariant and free of unphysical poles.

\subsection{Spin expansion and analysis of contact contributions}
Here we consider the expansion of the Compton amplitude reviewed above in powers of the spin vector and discuss the comparison between this Compton amplitude and known results in the literature computed from two different approaches, in particular, the higher-spin theory~\cite{Cangemi:2023bpe} and the Teukolsky equation~\cite{Bautista:2023sdf}. To avoid confusion, in the remaining part of this section, we refer to eq.~\eqref{amplitude structure} as $\cM_{\rm SP}$, where the subcript $\rm SP$ indicates that this expression is obtained from bootstrapping with only two parameters $x_1$ and $x_2$ and is speculated to describe the single-particle contributions only in~\cite{Bjerrum-Bohr:2023iey}.
Likewise, we denote the Compton amplitude computed from the higher-spin theory in~\cite{Cangemi:2023bpe} as $\cM_{\rm HS}$ and the one extracted from the solution to the Teukolsky equation in~\cite{Bautista:2023sdf} as $\cM_{\rm TS-FZ}$, where the ``FZ'' in the subscript indicates that we only consider the contributions obtained from imposing the so-called far-zone asymptotic behaviours.\footnote{We refer to the Teukolsky result with $\alpha\rightarrow 1$ as the ``far-zone data'' throught this manuscript, as indicated in~\cite{Bautista:2023sdf}. However, we acknowledge ambiguities in this way of separating the far-zone and near-zone contributions.}

The Compton amplitude derived from the higher-spin theory $\cM_{\rm HS}$ contains two entire functions which depend on three parameters~\cite{Cangemi:2023bpe}, among which is the spheroidicity parameter defined as $z = 2 \sqrt{-a\mdot a} (\sc p\mdot p_1)/m $~\cite{Dolan:2008kf} .\footnote{We use a different normalization for $z$ from that in~\cite{Cangemi:2023bpe}.} The $z$-dependence in $\cM_{\rm HS}$ follows from the higher-spin framework and shows up as contact terms only in the classical limit. Such terms are speculated to be related to the internal structures of the Kerr black hole~\cite{Kim:2023drc} or the near-zone physics, which is beyond the scope of $\cM_{\rm SP}$ given in eq.~\eqref{amplitude structure}.

On top of the discussions on the comparison at the level of the entire functions in~\cite{Bjerrum-Bohr:2023iey}, we have found further agreement with $\cM_{\rm HS}$ when taking $z\rightarrow 0$, namely
\begin{align}
   \cM_{\rm SP}(1^{\pm},2^{\mp},\sc p', \sc p)=\cM_{\rm HS}(1^{\pm},2^{\mp},\sc p', \sc p)|_{z\rightarrow 0}.
\end{align} 
This relation is checked order by order up to $\mathcal O(a^{20})$.

As for the comparison with~\cite{Bautista:2023sdf}, it is observed in~\cite{Bjerrum-Bohr:2023iey} that the extra contact terms needed for matching with the far-zone data at $\mathcal O(a^5)$ can be rewritten as a polynomial in $z$, although the $z$-dependence is traded for other parameters in the original expression of $\cM_{\rm TS-FZ}$. These extra contact terms vanish as $z\rightarrow 0$. Here we see that this observation holds up to $\mathcal O(a^8)$. In other words, at a given order in spin, we can separate $\cM_{\rm TS-FZ}$ into two parts, the $z$-independent part which agrees with $\cM_{\rm SP}$ in 4 dimensions and the $z$-dependent part which vanishes as $z\rightarrow 0$,
 \begin{align}
	\cM_{\rm SP}(1^{\pm},2^{\mp},\sc p', \sc p)= \cM_{\rm TS-FZ}(1^{\pm},2^{\mp},\sc p', \sc p)- \cM^{(c)}_{\rm TS-FZ} (z),\quad\text{with } \cM^{(c)}_{\rm TS-FZ} (0)=0.
\end{align}

$\cM_{\rm TS-FZ}$ is expressed in terms of the spinor-helicity variables. $\cM_{\rm SP}$ can be readily rewritten and it is straightforward to find $\cM_{\rm TS-FZ}^{(c)}$ in these variables.
But to see that $\cM_{\rm TS-FZ}^{(c)}$ indeed vanishes at a given order in spin as $z\rightarrow 0$, it is more convenient to use an alternative expression in terms of covariant variables. To this end, we adopt a simple procedure as follows. 

We begin with a general ansatz in terms of the covariant variables. That is, the ansatz is a linear combination of monomials comprised of powers of the parameter $z$ and factors of the forms $\mathsf v\mdot \mathsf v'$, $\mathsf v\mdot F_i\mdot F_j \mdot \mathsf v' $ and $\mathsf v\mdot F_i\mdot \Tilde{F}_j \mdot \mathsf v'$ where the vectors $\mathsf v$ and $\mathsf v'$ can be either $a^\mu$ or a momentum. The ansatz must have the correct counting in $a^\mu$ and $\varepsilon_i^\mu$ and satisfy the dimension and parity requirements. This naive ansatz contains 87 free parameters at $\mathcal O(a^5)$, 105 at $\mathcal O(a^6)$, 880 at $\mathcal O(a^7)$, and 540 at $\mathcal O(a^8)$. Odd orders in general have more structures to start with and thus there are seemingly more free parameters at $\mathcal O(a^7)$ than at $\mathcal O(a^8)$ in the first step. We can partially remove the redundancies due to the Levi-Civita symbols involved, by finding linear relations between the monomials in the ansatz. This leads to 9 parameters at $\mathcal O(a^5)$, 22 at $\mathcal O(a^6)$, 32 at $\mathcal O(a^7)$, and 55 at $\mathcal O(a^8)$. With the redundancies partially removed, we see that the number of free parameters grows as the order in spin increases.

Recall that the same-helicity gravitational Compton amplitude is known to all orders in spin, which is already captured in $\cM_{\rm SP}$. Demanding that $\cM_{\rm TS-FZ}^{(c)}(z)$ must vanish in the same-helicity configurations, we are left with 5 parameters at $\mathcal O(a^5)$, 12 at $\mathcal O(a^6)$, 17 at $\mathcal O(a^7)$, and 29 at $\mathcal O(a^8)$. Fitting the ansatz with the spinor-helicity expression of $\cM^{(c)}_{\rm TS-FZ}(z)$ in the opposite-helicity configurations, we find the covariant expression for $\cM_{\rm TS-FZ}^{(c)}$.

We note that the identities involving the Levi-Civita symbols only hold in 4 dimensions and may contain coefficients that are rational in inner products of momenta. Hence it is difficult to remove such redundancies in the ansatz completely. Consequently, there are a multitude of covariant expressions for $\cM_{\rm TS-FZ}^{(c)}$ at a given order in spin, although the expressions no longer contain explicit free parameters. They all evaluate to the same in 4 dimensions, but may be different in general dimensions. This leads to certain subtleties when we construct their contribution to one-loop amplitude, which will be discussed in the next section.

We give the explicit expressions for $\cM^{(c)}_{\rm TS-FZ}(z)$ at $\mathcal O(a^5)$ and $\mathcal O(a^6)$ and higher-order terms are given in Appendix \ref{app:Mc}. These expressions are manifestly gauge invariant and vanishing as $z\rightarrow 0$. We emphasize that they should be viewed as valid in 4 dimensions only, although they are expressed in terms of covariant quantities. 
\begin{align}\label{eq:explicitMc5Mc6}
\cM^{(c,5)}_{\rm TS-FZ}(z)&=2i p_1\mdot \bar p \, a\mdot a\left(a\mdot F_2\mdot \tF_1\mdot \bar p+a\mdot \tF_1\mdot F_2\mdot \bar p\right)\Bigg(\frac{\left(a\mdot a\right)  (\bar p\mdot F_1\mdot F_2\mdot \bar p) }{12 m^2}-\frac{11}{60}   (a\mdot F_1\mdot F_2\mdot a) \Bigg)\nn\\
    \cM^{(c,6)}_{\rm TS-FZ}(z)&=\frac{(a\mdot a)^3 \left(p_1\mdot \sc p\right){}^2 \left(\sc p\mdot F_1\mdot F_2\mdot \sc p\right){}^2}{9 m^4}+\frac{13 (a\mdot a)^3 p_1\mdot \sc p p_1\mdot F_2\mdot F_1\mdot \sc p \sc p\mdot F_1\mdot F_2\mdot \sc p}{90 m^2}\nn\\
    &-\frac{13 (a\mdot a)^3 p_1\mdot \sc p p_2\mdot F_1\mdot F_2\mdot \sc p \sc p\mdot F_1\mdot F_2\mdot \sc p}{90 m^2}+\frac{2 (a\mdot a)^2 p_1\mdot \sc p a\mdot p_1 \sc p\mdot F_1\mdot F_2\mdot \sc p a\mdot F_2\mdot F_1\mdot \sc p}{15 m^2}\nn\\
    &+\frac{16 (a\mdot a)^2 p_1\mdot \sc p a\mdot p_2 \sc p\mdot F_1\mdot F_2\mdot \sc p a\mdot F_2\mdot F_1\mdot \sc p}{45 m^2}-\frac{(a\mdot a)^2 \left(p_1\mdot \sc p\right){}^2 a\mdot F_1\mdot F_2\mdot a \sc p\mdot F_1\mdot F_2\mdot \sc p}{m^2}\nn\\
    &-\frac{2 (a\mdot a)^2 p_1\mdot \sc p a\mdot p_2 \sc p\mdot F_1\mdot F_2\mdot \sc p a\mdot F_1\mdot F_2\mdot \sc p}{15 m^2}-\frac{16 (a\mdot a)^2 p_1\mdot \sc p a\mdot p_1 \sc p\mdot F_1\mdot F_2\mdot \sc p a\mdot F_1\mdot F_2\mdot \sc p}{45 m^2}\nn\\
    &+\frac{14}{45} a\mdot a \left(p_1\mdot \sc p\right){}^2 \left(a\mdot F_1\mdot F_2\mdot a\right){}^2+\frac{22}{45} a\mdot a p_1\mdot \sc p a\mdot F_1\mdot F_2\mdot a a\mdot p_1 a\mdot F_1\mdot F_2\mdot \sc p\nn\\
    &-\frac{22}{45} a\mdot a p_1\mdot \sc p a\mdot F_1\mdot F_2\mdot a a\mdot p_2 a\mdot F_2\mdot F_1\mdot \sc p\,. 
\end{align}

Before closing this section, we make several remarks on the various attempts at obtaining the gravitational Compton in the literature. The Teukolsky solution yields the scattering phase of the graviton scattered off the Kerr black hole, by imposing the asymptotic behaviours of the Teukolsky equations in the near-zone (near the horizon) and the far-zone (at infinity)~\cite{Bautista:2022wjf,Bautista:2023szu,Bautista:2023sdf}. The scattering amplitude is then determined by matching at the level of the scattering phase. In principle, not only the leading but also higher-order contributions in the PM expansion can enter the non-perturbative scattering phase. Hence for a meaningful comparison with the tree Compton amplitudes obtained from field-theory approaches, it is crucial to isolate the leading PM contribution in the scattering phase. 
For a complete comparison between these two types of approaches, loop-level Compton amplitudes are also necessary.

On the other hand, the scattering phase obtained from the Teukolsky equation contains both near-zone and far-zone contributions. The far-zone contribution is rational, whereas the near-zone contains transcendental functions. The Compton amplitudes computed from field-theory approaches are postulated to capture the rational part. At the level of the scattering phase, there naturally is another type of ambiguities arising from rewriting the transcendental functions using various identities, which changes the rational terms. Therefore, it is also important to determine the rational terms in a way that is physically meaningful. 
In this paper, we consider and match with $\cM_{\rm TS-FZ}$ which is extracted from the far-zone contribution only. In \cite{Cangemi:2023bpe}, another procedure is discussed which leads to the matching between $\cM_{\rm HS}$ and $\cM_{\rm TS}$ at a different point. It would be interesting to see if there exists a way of rewriting the transcendental functions such that the remaining rational terms are independent of $z$.

\section{Eikonal with spin at 2PM}\label{sec:2PMAngle}
In this section, we compute the eikonal phase~\cite{Kosmopoulos:2021zoq} for the spinning Kerr binary at 2PM. The classical eikonal~\cite{Bern:2020buy} is an important generating that computes physical observables. In the non-spinning case, several variations, such as the radial action~\cite{Damgaard:2021ipf,Kol:2021jjc,Bjerrum-Bohr:2021wwt}, the HEFT phase~\cite{Brandhuber:2021eyq} and the WQFT eikonal~\cite{Mogull:2020sak,Wang:2022ntx,Jakobsen:2022psy}, have also been investigated. At 2PM, they are practically the equivalent, given by Fourier transforming the classical one-loop $2\rightarrow 2$ scattering amplitude to the impact parameter space. 
As seen in related works~\cite{Kosmopoulos:2021zoq,Witzany:2019nml}, these generating functions generally extend to the spinning case with spin-related subtleties. 

The eikonal and the resulting observables receive all types of contributions from the Compton amplitudes, as discussed in section~\ref{sec:ClassicalCompton}. 
Our main focus is the classical one-loop amplitude is constructed from the covariant Compton amplitude, using unitarity-based methods~\cite{Bjerrum-Bohr:2021vuf,Bjerrum-Bohr:2021din,Brandhuber:2021eyq}. We compute the general form of the one-loop integrand and perform the loop integration in the spin expansion explicitly up to $\mathcal O(a^8)$. 
For the sake of comparing with physical data, we include the contributions from the extra $z$-dependent contact terms at $\mathcal O(a^5)$ and $\mathcal O(a^6)$. Altogether, we find perfect agreement with the far-zone data. 

\subsection{Contribution from the covariant Compton amplitude}
The classical one-loop amplitude is constructed from the unitarity cut \cite{Bern:1994zx,Bern:1994cg,Bjerrum-Bohr:2013bxa,Cachazo:2017jef} and heavy-mass cut \cite{Brandhuber:2021eyq} as depicted below (with the mirror diagram implied), where the triple cut is given by two massless (graviton) cuts and one heavy-mass cut, 
\begin{align}
	\begin{tikzpicture}[baseline={([yshift=-0.8ex]current bounding box.center)}]\tikzstyle{every node}=[font=\small]	
\begin{feynman}
    	 \vertex (a) {\(\sc p_1\)};
    	 \vertex [right=1.5cm of a] (f2) [GR]{$~~\mathbf{S}~~$};
    	 \vertex [right=1.5cm of f2] (c){$\sc p'_1$};
    	 \vertex [above=2.0cm of a](ac){$\sc p_2$};
    	 \vertex [right=1.0cm of ac] (ad) [dot]{};
    	 \vertex [right=1.0cm of ad] (f2c) [dot]{};
    	  \vertex [above=2.0cm of c](cc){$\sc p'_2$};
    	  \vertex [above=1.0cm of a] (cutL);
    	  \vertex [right=3.0cm of cutL] (cutR);
    	  \vertex [right=0.5cm of ad] (att);
    	  \vertex [above=0.3cm of att] (cut20){$ $};
    	  \vertex [below=0.3cm of att] (cut21);
    	  \diagram* {
(a) -- [fermion,thick] (f2)-- [fermion,thick] (c),
    	  (f2)--[photon,ultra thick,momentum=\(\ell_1\)](ad), (f2)-- [photon,ultra thick,momentum'=\(\ell_2\)] (f2c),(ac) -- [fermion,thick] (ad)-- [fermion,thick] (f2c)-- [fermion,thick] (cc), (cutL)--[dashed, red,thick] (cutR), (cut20)--[ red,thick] (cut21)
    	  };
    \end{feynman}  
    \end{tikzpicture}
\quad\quad
   \begin{split}
   &     \mathcal M^{(1)}_{a_1 a_2} = {1\over 2}\sum_{h_i=\pm}{(32\pi G)^2 \over (4\pi)^{D/2}}\int {d^D \ell_1 \over \pi^{D/2}} {\delta ( m_2\, v_2\mdot\ell_1) \over \ell_1^2 \ell_2^2} \label{def:M1loop} \\
&\quad \;\times\Big(\mathcal M_3^{-h_1} (-\ell_1, v_2) \mathcal M_3^{-h_2} (-\ell_2, v_2) \mathcal M_4^{h_1 h_2} (\ell_1, \ell_2, v_1)\Big). 
    \end{split}
\end{align}
Here the massive cut labelled by a solid red line gives the $\delta$-function and we have formally lifted the massless cuts $\delta(\ell_j^2) \rightarrow i/\ell_j^2$, while keeping in mind that bubble or ultra-local terms, with zero or negative propagator powers for the massless propagators, should vanish. The velocities of the heavy bodies are parameterized such that $v_1^2 = v_2^2 =1$ and $v_1\mdot v_2 =\gamma$ as usual. The tree-level amplitudes $\cM_3$ and $\cM_4$ are given by eq.~\eqref{threepointgrav} and eq.~\eqref{amplitude structure}. We use dimensional regularization in the loop integration, namely $D=4-2\epsilon$. 
The summation is taken over the helicities of the gravitons and the completeness relation reads 
\begin{align}\label{completeness}
\sum_{h = \pm} \varepsilon^\mu_k \varepsilon^\nu_k \varepsilon^{*\rho}_{-k} \varepsilon^{*\sigma}_{-k}  = {1\over 2} \left(\mathcal P^{\mu\rho} \mathcal P^{\nu\sigma} + \mathcal P^{\mu\sigma} \mathcal P^{\nu\rho} \right) - {1 \over D-2} \mathcal P^{\mu\nu} \mathcal P^{\rho\sigma},
\end{align}
where
\begin{align}
\mathcal P^{\mu\nu} = \eta^{\mu\nu} - {k^\mu n^\nu + n^\mu k^\nu \over k\mdot n}\,
\end{align}
with the reference null vector $n^\mu$ transverse to the polarization $k\mdot\varepsilon = n\mdot\varepsilon=0$. The covariant three- and four-point tree amplitudes are expected to hold in general dimensions and in practice we can simply set $\mathcal P^{\mu\nu} = \eta^{\mu\nu}$. The reference vector always drops out. 

The dependence on the graviton polarization vectors in the tree amplitudes eq.~\eqref{threepointgrav} and eq.~\eqref{amplitude structure} always enters through the (dual) field strengths $F_i^{\mu\nu}$ and $\tilde F_i^{\mu\nu}$ and is completely manifest. 
Since the function $G_1(x)$ depends only on momenta and spins, the above gluing can be performed without expanding in powers of spin.
The resulting contribution to the one-loop integrand at all orders in spin is given in the \texttt{Ancillary Files}. 

The presence of $e^{\ell_1\mdot a_j}$ factors in eq.~\eqref{def:M1loop} may seem worrisome. Before expanding in powers of spin, potential divergences due to such factors can be easily remedies by a shift of the spin vector $a_j\rightarrow i\tilde a_j$. We illustrate with a simple example that the integral in eq.~\eqref{def:M1loop} does not behave worse than the non-spinning case does in the ultraviolet under this shift in Appendix \ref{app:UV}. Having evaluated the integral, we can analytically continue back to real values of $a_j$. As for the spin expansion of this integral, it can be defined as the Taylor expansion of the well-defined integral at each order. Similar integrals are studied and the resummed expressions are given in special kinematic regions in~\cite{Kim:2024grz}. Our preliminary analysis of simple examples in such kinematic regions show that the naive evaluation of the spin-expanded integral agrees with the expansion of the resummed one, order by order.

We now proceed to consider the expansion of eq.~\eqref{def:M1loop} in powers of spin. Expanding out the spin tensor $S^{\mu\nu}$ and using the identity\footnote{We use the package \texttt{KiHA} for manipulating the expressions~\cite{KiHA}.}
\begin{align}
    \epsilon^{\mu_1 \cdots \mu_n \alpha_1\cdots \alpha_m} \epsilon_{\mu_1\cdots \mu_n \beta_1\cdots \beta_m} = (-1)^n n! \delta^{\alpha_1\cdots \alpha_m}_{\beta_1\cdots \beta_m},\quad\quad 
    \delta^{\alpha_1\cdots \alpha_m}_{\beta_1\cdots \beta_m} = m!\delta^{\left[ \alpha_1 \right.}_{\; \beta_1} \cdots \delta^{\left. \alpha_m \right]}_{\;\beta_m},
\end{align}
we can always write the integrand in terms of scalar products and at most one factor of $\epsilon(A,B,C,D) = \epsilon_{\mu\nu\rho\sigma}A^\mu B^\nu C^\rho D^\sigma$. In fact, we find that at even orders in spin $\mathcal O(a^{2n})$, the integrand contains only scalar products and at odd orders in spin $\mathcal O(a^{2n+1})$, the integrand contains one $\epsilon(A,B,C,D)$ factor. Removing the bubble/ultra-local terms, the spin-expanded expression has two types of structures,
\begin{align}\label{eq:intTenStr}
\int {d^D \ell_1 \over \pi^{D/2}} {\delta (\ell_1\mdot v_2) (\ell_1\mdot a_1)^{n_1} (\ell_1\mdot a_2)^{n_2}   \over \ell_1^2 \ell_2^2 (\ell_1\mdot v_1)^m},\quad
\int {d^D \ell_1 \over \pi^{D/2}} {\delta (\ell_1\mdot v_2) (\ell_1\mdot a_1)^{n_1} (\ell_1\mdot a_2)^{n_2} \ell_1^\mu  \over \ell_1^2 \ell_2^2 (\ell_1\mdot v_1)^m},
\end{align}
where $n$, $n_1$, $n_2$ and $m$ are integers. 
The vector $\ell_1^\mu$ in the second term above typically gets contracted with $\epsilon(A,B,C,\cdot)_\mu$ where $A,B,C\in\{q,v_1,v_2,a_1,a_2\}$. 

Such one-loop tensor integrals can be readily decomposed into scalar integrals. These scalar integrals are dressed by rank-$n$ tensor structures. As shown in \cite{Jakobsen:2022fcj}, it is convenient to write down such tensor structures in terms of the following variables
\begin{align}\label{eq:tenVar}
    \theta^\mu_1 := {\gamma v_2^\mu-v_1^\mu \over \gamma^2 -1} , \quad
    \theta^\mu_2 := {\gamma v_1^\mu-v_2^\mu \over \gamma^2 -1} , \quad
    \Pi^{\mu\nu} := \eta^{\mu\nu}-\theta_1^\mu v_1^\nu - \theta_2^\mu v_2^\nu - {q^\mu q^\nu \over q^2},
\end{align}
where $\Pi^{\mu\nu}$ is the $(D-3)$-dimensional metric. It is easy to verify that a tensor integral of the form 
\begin{align}
\mathcal I_{1,1,m,1}[\ell_1^{\mu_1}  \cdots \ell_1^{\mu_n}] =  \int {d^D \ell_1 \over \pi^{D/2}} {\delta (\ell_1\mdot v_2) \ell_1^{\mu_1} \cdots \ell_1^{\mu_n}  \over \ell_1^2 \ell_2^2 (\ell_1\mdot v_1)^m}
\end{align}
can be decomposed into scalar integrals dressed by tensor structures built from $q^\mu$, $\theta_1$ and $\Pi^{\mu\nu}$ only. Since all but at most one of these indices are to be contracted with $a_1$ or $a_2$, symmetries of the two structures in eq.~\eqref{eq:intTenStr} lead to closed-form expressions for the decomposition. We present the explicit decomposition of the first structure in eq.~\eqref{eq:intTenStr} here and postpone the derivation to Appendix \ref{app:TenDec},
\begin{align}
&\mathcal I_{1,1,\alpha,1}[(\ell_1\mdot a_1)^M (\ell_1\mdot a_2)^{N-M}] \nn \\
= & \sum_{\substack{n_1+n_2+2n_3=N, \\ n_i\geqslant 0, n_i\in\mathbb Z}} {1\over N_{n_3}} \left( \sum_{k=0}^{n_3} \left(\begin{array}{c}
         n_3  \\
          k
    \end{array}\right) {1\over (\gamma^2 -1)^k} \left({q^2 \over 2}\right)^{n_2+n_3-k} \mathcal I_{1,1,\alpha-n_1-2k,1}[1] \right) \nn \\
    & \quad\quad\times \Bigg( \sum_{\text{cond.}} C_{N,M,n_1,n_2,m_1,m_2,m_3}~ (a_1\mdot\theta_1)^{m_1} (a_2\mdot\theta_1)^{n_1-m_1} (a_1\mdot q)^{m_2} (a_2\mdot q)^{n_2-m_2} \nn \\
   & \quad\quad\quad\quad\quad\quad  (a_1\mdot\Pi \mdot a_1)^{m_3} (a_1\mdot\Pi\mdot a_2)^{m_4} (a_2\mdot\Pi\mdot a_2)^{n_3-m_3-m_4} \Bigg),
\end{align}
where $N_{n_3} = (D-3) (D-1) \cdots (D-3+2(n_3-1))$ for $n_3 >0$ and $N_0 =1$. The summation in the last bracket is taken over all the solutions to the conditions below
\begin{align}
    m_1+m_2+2m_3+m_4 =M,\quad
    0\leqslant m_i \leqslant n_i, \quad 0\leqslant m_4 \leqslant n_3, 
    \quad m_i\in\mathbb Z.
\end{align}
The coefficients read
\begin{align}
    C_{N,M,n_1,n_2,m_1,m_2,m_3} = { M! \over m_1! m_2! m_3! m_4!} {(N-M)! \over (n_1-m_1)! (n_2-m_2)! (n_3-m_3)!} {1\over 2^{n_3-m_4}} .
\end{align}
The second structure in eq.~\eqref{eq:intTenStr} admits a similar expression with one free index, which we also give in Appendix \ref{app:TenDec}.
 
After decomposing the tensor integrals, the resulting scalar integrals are easily cast in a basis of master integrals using the IBP relations.\footnote{We use the package \texttt{LiteRed} for the IBP reduction \cite{Lee:2012cn,Lee:2013mka}.} Up to $\mathcal O(a^8)$, we see that only the triangle integral below contributes in 4 dimensions,\footnote{The box integral $\mathcal I_{1,1,1,1}$ appears after the IBP reduction. But its coefficient vanishes when we restrict to 4 dimensions.}
\begin{align}
\mathcal I_{1,1,0,1} = \int {d^D \ell_1 \over \pi^{D/2}} {\delta (\ell_1\mdot v_2) \over \ell_1^2 \ell_2^2} = {2^{5-D} \pi^2 (-q^2)^{(D-5)/2}~ \text{sec}\left( \pi D/2 \right) \over \Gamma\left( {D\over 2}-1 \right) }.
\end{align}

Evaluating the integral, we arrive at the spin-expanded one-loop amplitude computed from the covariant Compton amplitude in eq.~\eqref{amplitude structure}. It is independent of the $z$-parameter.   
We find perfect agreements with the literature up to $\mathcal O(G^2 a^4)$~\cite{Bautista:2023szu}. Since the terms at low orders in spin are well established, we choose not to display the relatively bulky expressions. Starting from $\mathcal O(G^2 a^5)$, to match with the far-zone data, $z$-dependent contact terms $\cM_{\rm TS-FZ}^{(c)}(z)$ need to be included. The gluing process involving eq.~\eqref{eq:explicitMc5Mc6} is slightly more subtle, which we are to discuss shortly.
We include the spin-expanded one-loop amplitude from $\mathcal O(a^5)$ to $\mathcal O(a^8)$ in the \texttt{Ancillary Files}.

\subsection{Contribution from extra contact terms}
Here we discuss the one-loop contribution from the $z$-dependent contact terms $\cM_{\rm TS-FZ}^{(c)}$ at $\mathcal O(G^2 a^5)$ and $\mathcal O(G^2 a^6)$.  It is given by the diagram depicted in eq.~\eqref{def:M1loop} with the four-point Compton amplitude substituted by eq.~\eqref{eq:explicitMc5Mc6}. Since the covariant Compton amplitude in~eq.~\eqref{amplitude structure} suffices to match with the far-zone data up to $\mathcal O(a^4)$, the $\mathcal O(G^2 a^5)$ $z$-dependence only enters through the $\mathcal O(G^2 a_1^5 a_2^0)$ or $\mathcal O(G^2 a_1^0 a_2^5)$ terms. Hence the three-point amplitude can be taken as the non-spinning one. At $\mathcal O(G^2 a^6)$, we need both $\mathcal O(G^2 a^5_1 a_2^1)$ and $\mathcal O(G^2 a_1^6 a_2^0)$ (and their mirrors), which come from gluing up the two expressions in eq.~\eqref{eq:explicitMc5Mc6} with appropriate orders in the spin expansion of the three-point amplitude respectively. 

As demonstrated in the previous section, the $z$-dependent contact terms in eq.~\eqref{eq:explicitMc5Mc6} are constructed such that they agree with those in~\cite{Bautista:2023sdf} in 4 dimensions. This allows for ambiguities and the covariant expressions, which are equivalent in 4 dimensions may have subtle differences, if naively extrapolated to general dimensions. Hence, their contribution to the one-loop amplitude needs to be constructed strictly in 4 dimensions as well. 

In 4 dimensions, the three-point amplitude in eq.~\eqref{threepointgrav} simplifies to
\begin{align}
\mathcal M_3(1^+,\bar p', \bar p) = -i\kappa e^{+p_1\mdot a} (\bar p\mdot\varepsilon_1^+)^2,\quad\quad
\mathcal M_3(1^-,\bar p', \bar p) = -i\kappa e^{-p_1\mdot a} (\bar p\mdot\varepsilon_1^-)^2.
\end{align}
Recall that the extra contact terms $\cM_{\rm TS-FZ}^{(c)}$ also vanish in the same-helicity configurations, namely, $\cM_{\rm TS-FZ}^{(c,5)}(1^\pm,2^\pm,\bar p',\bar p)=\cM_{\rm TS-FZ}^{(c,6)}(1^\pm,2^\pm,\bar p',\bar p) =0$. Consequently, their one-loop contribution is given by
\begin{align}
\!\!\!\! \begin{tikzpicture}[baseline={([yshift=-0.8ex]current bounding box.center)}]\tikzstyle{every node}=[font=\small]	
\begin{feynman}
    	 \vertex (a) {\(\sc p_1\)};
    	 \vertex [right=1.5cm of a] (f2) [GR]{$~~~$};
    	 \vertex [right=1.5cm of f2] (c){$\sc p'_1$};
    	 \vertex [above=2.0cm of a](ac){$\sc p_2$};
    	 \vertex [right=1.0cm of ac] (ad) [dot]{};
    	 \vertex [right=1.0cm of ad] (f2c) [dot]{};
    	  \vertex [above=2.0cm of c](cc){$\sc p'_2$};
    	  \vertex [above=1.0cm of a] (cutL);
    	  \vertex [right=3.0cm of cutL] (cutR);
    	  \vertex [right=0.5cm of ad] (att);
    	  \vertex [above=0.3cm of att] (cut20){$ $};
    	  \vertex [below=0.3cm of att] (cut21);
         \vertex [above right=0.2cm and 0.45cm of cutL] (hpL) {$+$};
         \vertex [below right=0.2cm and 0.6cm of cutL] (hmL) {$-$};
        \vertex [above left=0.2cm and 0.45cm of cutR] (hmR) {$-$};
         \vertex [below left=0.2cm and 0.6cm of cutR] (hpR) {$+$};
    	  \diagram* {
(a) -- [fermion,thick] (f2)-- [fermion,thick] (c),
    	  (f2)--[photon,ultra thick](ad), (f2)-- [photon,ultra thick] (f2c),(ac) -- [fermion,thick] (ad)-- [fermion,thick] (f2c)-- [fermion,thick] (cc), (cutL)--[dashed, red,thick] (cutR), (cut20)--[ red,thick] (cut21)
    	  };
    \end{feynman}  
    \end{tikzpicture} 
   + 
   \begin{tikzpicture}[baseline={([yshift=-0.8ex]current bounding box.center)}]\tikzstyle{every node}=[font=\small]	
\begin{feynman}
    	 \vertex (a) {\(\sc p_1\)};
    	 \vertex [right=1.5cm of a] (f2) [GR]{$~~~$};
    	 \vertex [right=1.5cm of f2] (c){$\sc p'_1$};
    	 \vertex [above=2.0cm of a](ac){$\sc p_2$};
    	 \vertex [right=1.0cm of ac] (ad) [dot]{};
    	 \vertex [right=1.0cm of ad] (f2c) [dot]{};
    	  \vertex [above=2.0cm of c](cc){$\sc p'_2$};
    	  \vertex [above=1.0cm of a] (cutL);
    	  \vertex [right=3.0cm of cutL] (cutR);
    	  \vertex [right=0.5cm of ad] (att);
    	  \vertex [above=0.3cm of att] (cut20){$ $};
    	  \vertex [below=0.3cm of att] (cut21);
         \vertex [above right=0.2cm and 0.45cm of cutL] (hpL) {$-$};
         \vertex [below right=0.2cm and 0.6cm of cutL] (hmL) {$+$};
        \vertex [above left=0.2cm and 0.45cm of cutR] (hmR) {$+$};
         \vertex [below left=0.2cm and 0.6cm of cutR] (hpR) {$-$};
    	  \diagram* {
(a) -- [fermion,thick] (f2)-- [fermion,thick] (c),
    	  (f2)--[photon,ultra thick](ad), (f2)-- [photon,ultra thick] (f2c),(ac) -- [fermion,thick] (ad)-- [fermion,thick] (f2c)-- [fermion,thick] (cc), (cutL)--[dashed, red,thick] (cutR), (cut20)--[ red,thick] (cut21)
    	  };
    \end{feynman}  
    \end{tikzpicture}
    =
    \begin{split}
     & {1\over 2} {(32\pi G)^2 \over (4\pi)^{D/2}}\int {d^D \ell_1 \over \pi^{D/2}} {\delta ( m_2\, v_2\mdot\ell_1) \over \ell_1^2 \ell_2^2} \\
    &\Big( e^{(\ell_1-\ell_2)\mdot a_2} \mathcal C_{+-} + e^{-(\ell_1-\ell_2)\mdot a_2} \mathcal C_{-+} \Big),
    \end{split}
\end{align}
where
\begin{align}
    \mathcal C_{\pm \mp} = \left.(\bar p_2\mdot\varepsilon_1^\pm)^2 (\bar p_2\mdot\varepsilon_2^\mp)^2 \cM_{\rm TS-FZ}^{(c)}(\ell_1^\mp,\ell_2^\pm,\bar p'_1,\bar p_1) \right|_{\text{ completeness relation}}
\end{align}
In this case, we have to use the ``chiral'' version of the completeness relation. Equivalently, we can separate the even and odd terms in spin, as discussed in~\cite{Brandhuber:2024bnz}. At even orders in spin, we have
\begin{align}
    \cosh\left(\ell_{12}\mdot a_2\right)(\mathcal C_{+-}+\mathcal C_{-+}) = \cosh\left(\ell_{12}\mdot a_2\right) (\mathcal C_{+-}+\mathcal C_{++}+\mathcal C_{--}+\mathcal C_{-+}),
\end{align}
where $\ell_{12}=\ell_1-\ell_2$ and $\mathcal C_{++}+\mathcal C_{--}=0$. Here we can use the same completeness relation in eq.~\eqref{completeness}, since we have
\begin{align}
    \varepsilon^{+\mu}_k \varepsilon^{+\nu}_k\varepsilon^{-\rho}_{-k} \varepsilon^{-\sigma}_{-k} +    \varepsilon^{-\mu}_k \varepsilon^{-\nu}_k\varepsilon^{+\rho}_{-k} \varepsilon^{+\sigma}_{-k} 
\end{align}
for both gravitons. 
At odd powers in spin, we have instead
\begin{align}
    \sinh\left(\ell_{12}\mdot a_2\right)(\mathcal C_{+-}-\mathcal C_{-+}) = \sinh\left(\ell_{12}\mdot a_2\right) (\mathcal C_{+-}-\mathcal C_{++}+\mathcal C_{--}-\mathcal C_{-+}).
\end{align}
Hence we have 
\begin{align}
        \varepsilon^{+\mu}_k \varepsilon^{+\nu}_k\varepsilon^{-\rho}_{-k} \varepsilon^{-\sigma}_{-k} -    \varepsilon^{-\mu}_k \varepsilon^{-\nu}_k\varepsilon^{+\rho}_{-k} \varepsilon^{+\sigma}_{-k} = {1\over 4} \left(\mathcal P^{\mu\rho}\tilde{\mathcal P}^{\nu\sigma} + \mathcal P^{\mu\sigma}\tilde{\mathcal P}^{\nu\rho} +\mathcal P^{\nu\rho}\tilde{\mathcal P}^{\mu\sigma} + \mathcal P^{\nu\sigma}\tilde{\mathcal P}^{\mu\rho}    \right),
\end{align}
for one of the two gravitons, where
\begin{align}
    \tilde{\mathcal P}^{\mu\nu} = {i\epsilon^{\mu\nu\rho\sigma} k_\rho n_\sigma \over k\mdot n}.
\end{align}
This way we guarantee that the remnant ambiguities introduced in the process of lifting the spinor-helicity expression for $\cM_{\rm TS-FZ}^{(c)}$ to the covariant one have no impact at one loop. 

This concludes our discussion on the one-loop contribution from the extra contact terms. Together with the contribution from the covariant Compton amplitude, we find agreement with the far-zone data in~\cite{Bautista:2023szu} up to $\mathcal O(G^2 a^6)$, which is the highest-order result to our best knowledge. Comparing with the 2PM calculation from the higher-spin amplitudes~\cite{Bohnenblust:2024hkw} may be helpful for finding an all-order-in-spin remedy for the $z$-dependence.

\subsection{2PM eikonal in the spin expansion}
Generally speaking, the eikonal exponentiation can be identified with the classical amplitude Fourier transformed to the impact parameter space. The classical eikonal phase is then extracted order by order. Recent works~\cite{Kosmopoulos:2021zoq,Kim:2024grz} show that the classical eikonal can be viewed as a generator for canonical transformations.\footnote{Another work~\cite{Gonzo:2024zxo} shows that the radial action can be understood in a similar fashion.}

Up to 2PM, the relation between the amplitude and the eikonal remains straightforward, namely
\begin{align}
    \chi^{\rm (nPM)}_\tau = {1\over 4 m_1 m_2 \sqrt{\gamma^2-1}}\int {d^{D-2} \mathbf{q} \over (2\pi)^{D-2}} e^{-i\mathbf{b}\mdot \mathbf{q}} \cM^{(n-1)}_{a_1a_2} U_1 U_2,
\end{align}
where $\mathbf{b}$ is the impact parameter and $\tau$ labels the Thomas-Wigner rotation factors~\cite{Chung:2019duq,Chung:2020rrz} 
\begin{align}
    U_i = e^{i \tau \mathcal E_i \over E m_i (E_i + m_i)},\quad\quad \mathcal E_i := E \, \mathbf{s}_i\mdot (\mathbf{p}\times \mathbf{q})
\end{align}
with $E_i$ denoting the energy of the two black holes and $E=E_1+E_2$. The bold-faced letters denote the three-vectors. The three-momenta are given in the center-of-mass (COM) frame, namely $p_1 = (E_1,\mathbf{p}),~ p_2=(E_2,-\mathbf{p}),~q=(0,\mathbf{q})$, and $\mathbf{s}_i$ is the spatial part of $s_i^\mu = m_i a_i^\mu$. Different values of $\tau$ correspond to different choices of the spin supplementary conditions (SSC): $\tau=0$ corresponds to the covariant SSC $\bar{p}_\mu S^{\mu\nu}=0$ whereas $\tau=1$ the canonical SSC. (For more details, see also \cite{Chung:2018kqs,Bautista:2023szu}.) Dressing the amplitude with the Thomas-Wigner rotation factors is effectively equivalent to shifting the impact parameter $\mathbf{b}\rightarrow \mathbf{b} + \sum_{i=1,2}\left({-\tau \over E_i+m_i}\right) \mathbf{p}\times \mathbf{s}_i$. The canonical observables follow from the canonical SSC.

Here we showcase the explicit expression for the $z$-independent $\mathcal O(G^2 a^7)$ contributions to $\chi^{\rm (2PM)}_{\tau=0}$ in the aligned-spin case, which is given by the covariant Compton amplitude. To compare with the far-zone data, it needs to be supplemented by the contributions from the $z$-dependent contact terms $\cM_{\rm TS-FZ}^{(c,n=5,6,7)}$, which are given in eq.~\eqref{eq:explicitMc5Mc6} and Appendix~\ref{app:Mc}. We parameterize the generalised aliged spins as $a_2^\mu = \xi a_1^\mu$ and hence the powers of $a_1$ and $a_2$ are tracked by the power of $\xi$. 
\begin{align}
\left. \chi^{\rm (2PM)}_{\tau=0} \right|_{\xi^4, z=0} &= -\frac{35 \pi  G^2 \xi ^4 \gamma \left(24 \left(9 \gamma ^2-4\right) m_1+ \left(114-239 \gamma ^2\right) m_2\right) }{512 \left(\gamma
   ^2-1\right)^{3/2} |\mathbf{b}|^8}~ \hat{b}\mdot S_1\mdot p_2 \nn\\
&\quad\quad    \left({429
   (\hat{b}\mdot a_1 )^6}- {594 a_1^2 (\hat{b}\mdot a_1 )^4}+{216 a_1^4
   ( \hat{b} \mdot a_1)^2 }-16 a_1^6 \right),  \\
\left. \chi^{\rm (2PM)}_{\tau=0} \right|_{\xi^5, z=0} &= -\frac{21 \pi  G^2 \xi ^5 \gamma \, \hat{b}\mdot S_1\mdot p_2}{512 \left(\gamma ^2-1\right)^{3/2}
  |\mathbf{b}|^8}  \Big(2145 \left(\left(38 \gamma ^2-15\right) m_1+26 \left(1-2 \gamma ^2\right)
   m_2\right) (\hat{b}\mdot a_1)^6 \nn \\
&  \quad\quad\quad -33 a_1^2  \big(\left(3421 \gamma ^2-1351\right) m_1+2340 \left(1-2 \gamma
   ^2\right) m_2\big) (\hat{b}\mdot a_1)^4 \nn\\
&  \quad\quad\quad  +36 a_1^4  \big(\left(1141 \gamma ^2-451\right)
   m_1+780 \left(1-2 \gamma ^2\right) m_2\big) (\hat{b}\mdot a_1)^2 \nn\\
&  \quad\quad\quad  -8 a_1^6 
   \big(\left(381 \gamma ^2-151\right) m_1+260 \left(1-2 \gamma ^2\right) m_2\big)\Big), \\
\left. \chi^{\rm (2PM)}_{\tau=0} \right|_{\xi^6, z=0} &= -\frac{7 \pi   G^2 \xi ^6 \gamma \, \hat{b}\cdot S_1\cdot p_2}{4096
   \left(\gamma ^2-1\right)^{3/2} |\mathbf{b}|^8} \Big(\!\!-\!\!16 a_1^6  \big(16 \left(81 \gamma ^2-26\right) m_1+15 \left(77-149 \gamma ^2\right) m_2\big) \nn\\
 &\quad\quad\quad  -66 a_1^2  \big(16 \left(721 \gamma ^2-226\right) m_1+135
   \left(77-149 \gamma ^2\right) m_2\big) (\hat{b}\mdot a_1)^4 \nn\\
 &\quad\quad\quad  +72 a_1^4  \big(16
   \left(241 \gamma ^2-76\right) m_1+45 \left(77-149 \gamma ^2\right) m_2\big) (\hat{b}\mdot a_1)^2\nn\\
&\quad\quad\quad   + 2145 \left(16 \left(16 \gamma ^2-5\right) m_1+3 \left(77-149 \gamma
   ^2\right) m_2\right) (\hat{b}\mdot a_1)^6 \Big), \\
\left. \chi^{\rm (2PM)}_{\tau=0} \right|_{\xi^7, z=0} &= -\frac{\pi  G^2 \xi ^7  \gamma \,\hat{b}\mdot S_1\mdot p_2 }{1024 \left(\gamma ^2-1\right)^{3/2}
   |\mathbf{b}|^8}\Big(2145 \left(6 \left(8 \gamma ^2-1\right) m_1+7 \left(9-17 \gamma ^2\right)
   m_2\right) (\hat{b}\mdot a_1)^6 \nn \\
&\quad\quad\quad   -198 a_1^2 \big(\left(727 \gamma ^2-97\right) m_1+105 \left(9-17 \gamma
   ^2\right) m_2\big) (\hat{b}\mdot a_1)^4 \nn\\
&\quad\quad\quad   +108 a_1^4  \big(\left(493 \gamma ^2-73\right)
   m_1+70 \left(9-17 \gamma ^2\right) m_2\big) (\hat{b}\mdot a_1)^2\nn\\
&\quad\quad\quad   -8 a_1^6 \big(\left(513 \gamma ^2-93\right) m_1+70 \left(9-17 \gamma ^2\right) m_2\big)\Big),
\end{align}
where we define $\hat{b}^\mu = b^\mu /|b|$ and in this configuration $\hat{b}\mdot a_1 \neq 0$. The terms that are not listed above can be obtained by swapping the two particles $\chi^{\rm (2PM)}|_{\xi^n} = \left(\chi^{\rm (2PM)}|_{\xi^{7-n}}\right)_{1\leftrightarrow 2}$. 

Restricting to the aligned-spin configuration ($a_1^\mu \parallel a_2^\mu$ and $b\cdot a_i$=0), it is straightforward to obtain the scattering angle from the eikonal under the covariant SSC via
\begin{align}
    \theta_{\rm (nPM)} = {E \over \sqrt{\gamma^2-1}} {\partial \over \partial |\mathbf{b}|} \chi^{\rm (nPM)}_{\tau=0}\,.
\end{align}
Hence at the seventh power in spin, the $z$-independent contribution to the 2PM scattering angle in the aligned-spin case reads,
\begin{align}
    \theta_{\rm (2PM), z=0} = & \frac{\pi G^2 E\, \xi ^4 \gamma\, a_1^6\, \hat{b}\mdot S_1\mdot p_2}{32 \left(\gamma ^2-1\right)^2 |\mathbf{b}|^9} \Bigg( m_1\Big(6 \left(171 \gamma ^2-31\right) \xi ^3+112 \left(81 \gamma ^2-26\right) \xi ^2 \nn\\
  &  +84 \left(381 \gamma ^2-151\right) \xi +6720 \left(9
   \gamma ^2-4\right)\Big) + m_2 \Big( 140 \left(9-17 \gamma ^2\right) \xi ^3  \nn\\
   &  +\left(8085-15645 \gamma ^2\right) \xi ^2-21840 \left(2 \gamma ^2-1\right) \xi -66920 \gamma
   ^2+31920 \Big)
   \Bigg)\nn\\
  & + \left( m_1\leftrightarrow m_2 \,, \xi^n \leftrightarrow \xi^{7-n} \right)\,.
\end{align}
We note that the $\mathcal O(\xi^3)$ and $\mathcal O(\xi^4)$ terms in the above expression agree with those in~\cite{Bohnenblust:2024hkw} while the differences in other terms are accounted for by the contributions from the $z$-dependent contact terms given in \eqref{eq:explicitMc5Mc6} and Appendix~\ref{app:Mc}.

\section{Conclusions and outlook}
We consider a particular proposal for the classical gravitational Compton amplitude, which follows from bootstrapping techniques and is expressed in covariant variables. This covariant expression is consistent with all possible massless and massive factorization requirements at all orders in spin and contains a contact contribution that is obtained by imposing the same empirical patterns as those observed in other contributions.

In this paper, we further the analysis in~\cite{Bjerrum-Bohr:2023iey} regarding the contact terms. In particular, we verify the matching between the covariant Compton amplitude and the one derived from the higher-spin theory~\cite{Cangemi:2023bpe} up to $\mathcal O(a^{20})$, when taking the spheroidicity parameter $z\rightarrow 0$. We also devise a simple procedure to find a covariant form which accounts for the missing contact contributions compared to those extracted from the far-zone data. We confirm that such contributions can be written as a polynomial in $z$ and vanish as $z\rightarrow 0$ up to $\mathcal O(a^8)$. We believe such patterns extend to higher orders in spin.
These observations corroborate somewhat the folklore that the $z$-dependent contact terms are associated with the internal structures of the Kerr black hole. Further studies on multiple fronts are needed to uncover the precise nature of such contact contributions.  

From the covariant Compton amplitude, the calculation of the classical one-loop amplitude on the triple cut using unitarity-based methods is then streamlined. 
We obtain an all-order-in-spin integrand and evaluate the integral up to $\mathcal O(a^8)$ in the spin expansion. We have also included the contributions from $z$-dependent contact terms, for which the computation involves an extra subtlety. The Fourier transform of the one-loop amplitude to the impact parameter space gives the eikonal phase. We find perfect agreement with the 2PM far-zone data up to $\mathcal O(a^6)$.
It is certainly interesting to consider the one-loop integral without expanding in spin. 
Two directions are natural to look into, the binary dynamics and the waveform at higher PM orders. The binary dynamics at $\mathcal O(G^3)$ is given by the two-loop $2\rightarrow 2$ amplitude and the waveform at $\mathcal O(G^{5/2})$ is given by the one-loop amplitude with a graviton radiation. The five-point tree-level gravitational Compton amplitude is needed in both cases. Computing the five-point amplitude to an arbitrary order in spin is by itself a challenge, although it is conceivable that the bootstrapping techniques should carry over. Even without the five-point amplitude at hand, one may consider mass sectors, for instance the ``zig-zag'' diagrams at two loops in~\cite{Brandhuber:2021eyq}, which only involve the three- and four-point amplitudes in the construction of loop amplitudes. The behaviours of the loop amplitudes in these sectors may in turn shed new light on additional constraints for the structures of the gravitational Compton amplitudes.

\section*{Acknowledgement}
We thank F. Y. Baustista, E. Bjerrum-Bohr, A. Brandhuber, G. R. Brown, J. Gowdy,  H. Johansson,  J.-W. Kim,  S. Lee, P. Pichini,  M. Skowronek, G. Travaglini for insightful discussions. TW is grateful to H. Johansson, J.-W. Kim and S. Lee for discussions on their respective works and comments on the manuscript. GC has received funding from the European Union's Horizon 2020 research and innovation program under the Marie Sk\l{}odowska-Curie grant agreement No.~847523 ``INTERACTIONS''. TW is supported by the NRF grant 2021R1A2C2012350 and the Fellowship of China Postdoctoral Science Foundation (No. 2022M713228).

\appendix
\section{Extra contact terms}\label{app:Mc}
Here we present the explicit expressions for the extra contact terms that give the $z$-dependence of the Compton amplitude extracted from the Teukolsky solution at $\mathcal O(a^7)$ and $\mathcal O(a^8)$. We choose only the contribution obtained from fitting the far-zone asymptotic. 
\begin{align}
    &\cM^{(c,7)}_{\rm TS-FZ}= \frac{\left(284 a\mdot p_1+211 a\mdot p_2\right) \left(p_1\mdot \sc p\right){}^2 \sc p\mdot F_1\mdot F_2\mdot \sc p \sc p\mdot F_1\mdot \tF_2\mdot \sc p (a\mdot a)^3}{945 m^4}\nn\\
    &+\frac{8 p_1\mdot \sc p \left(a\mdot F_1\mdot p_2 p_1\mdot F_2\mdot \sc p+a\mdot F_2\mdot p_1 p_2\mdot F_1\mdot \sc p\right) \left(\sc p\mdot F_1\mdot \tF_2\mdot \sc p-\sc p\mdot F_2\mdot \tF_1\mdot \sc p\right) (a\mdot a)^3}{945 m^2}\nn\\
    &+\frac{\left(211 a\mdot p_1+284 a\mdot p_2\right) \left(p_1\mdot \sc p\right){}^2 \sc p\mdot F_1\mdot F_2\mdot \sc p \sc p\mdot F_2\mdot \tF_1\mdot \sc p (a\mdot a)^3}{945 m^4}\nn\\
    &+\frac{p_1\mdot \sc p \left(99 \left(a\mdot p_1\right){}^2 m^2-101 a\mdot p_1 a\mdot p_2 m^2-78 a\mdot a \left(p_1\mdot \sc p\right){}^2\right) a\mdot F_1\mdot F_2\mdot \sc p \sc p\mdot F_1\mdot \tF_2\mdot \sc p (a\mdot a)^2}{1890 m^4}\nn\\
    &+\frac{4 p_1\mdot \sc p \left(-33 \left(a\mdot p_1\right){}^2 m^2+23 a\mdot p_1 a\mdot p_2 m^2+60 a\mdot a \left(p_1\mdot \sc p\right){}^2\right) a\mdot F_2\mdot F_1\mdot \sc p \sc p\mdot F_1\mdot \tF_2\mdot \sc p (a\mdot a)^2}{945 m^4}\nn\\
    &+\frac{p_1\mdot \sc p \left(-99 \left(a\mdot p_2\right){}^2 m^2+101 a\mdot p_1 a\mdot p_2 m^2+78 a\mdot a \left(p_1\mdot \sc p\right){}^2\right) a\mdot F_2\mdot F_1\mdot \sc p \sc p\mdot F_2\mdot \tF_1\mdot \sc p (a\mdot a)^2}{1890 m^4}\nn\\
    &-\frac{101 a\mdot p_1 \left(p_1\mdot \sc p\right){}^2 a\mdot F_1\mdot F_2\mdot a \sc p\mdot F_1\mdot \tF_2\mdot \sc p (a\mdot a)^2}{135 m^2}-\frac{101 a\mdot p_2 \left(p_1\mdot \sc p\right){}^2 a\mdot F_1\mdot F_2\mdot a \sc p\mdot F_2\mdot \tF_1\mdot \sc p (a\mdot a)^2}{135 m^2}\nn\\
    &-\frac{4 p_1\mdot \sc p \left(-33 \left(a\mdot p_2\right){}^2 m^2+23 a\mdot p_1 a\mdot p_2 m^2+60 a\mdot a \left(p_1\mdot \sc p\right){}^2\right) a\mdot F_1\mdot F_2\mdot \sc p \sc p\mdot F_2\mdot \tF_1\mdot \sc p (a\mdot a)^2}{945 m^4}\nn\\
    &+\frac{212}{945} a\mdot p_1 \left(p_1\mdot \sc p\right){}^2 a\mdot F_1\mdot F_2\mdot a a\mdot F_1\mdot \tF_2\mdot a a\mdot a+\frac{212}{945} a\mdot p_2 \left(p_1\mdot \sc p\right){}^2 a\mdot F_1\mdot F_2\mdot a a\mdot F_2\mdot \tF_1\mdot a a\mdot a\nn\\
    &+\frac{p_1\mdot \sc p \left(-121 \left(a\mdot p_1\right){}^2 m^2-121 \left(a\mdot p_2\right){}^2 m^2+190 a\mdot p_1 a\mdot p_2 m^2+372 a\mdot a \left(p_1\mdot \sc p\right){}^2\right) }{630 m^2}\nn\\
    &~~~\times ( a\mdot F_1\mdot F_2\mdot a a\mdot \tF_1\mdot F_2\mdot \sc p a\mdot a)
\end{align}
\begin{align}
   &\cM^{(c,8)}_{\rm TS-FZ}= \frac{64 \left(p_1\mdot \sc p\right){}^2 p_1\mdot F_2\mdot \sc p p_2\mdot F_1\mdot \sc p \sc p\mdot F_1\mdot F_2\mdot \sc p (a\mdot a)^4}{2835 m^4}+\frac{32 \left(p_1\mdot \sc p\right){}^2 p_1\mdot F_2\mdot F_1\mdot p_2 \sc p\mdot F_1\mdot F_2\mdot \sc p (a\mdot a)^4}{2835 m^2}\nn\\
   &+\frac{\left(p_1\mdot \sc p\right){}^2 \left(m^2 \left(174 \left(a\mdot p_1\right){}^2-55 a\mdot p_2 a\mdot p_1+174 \left(a\mdot p_2\right){}^2\right)-278 a\mdot a \left(p_1\mdot \sc p\right){}^2\right) \left(\sc p\mdot F_1\mdot F_2\mdot \sc p\right){}^2 (a\mdot a)^3}{5670 m^6}\nn\\
   &+\frac{8 \left(a\mdot p_1-a\mdot p_2\right)}{945 m^2} \Big(p_1\mdot \sc p a\mdot F_1\mdot p_2 p_1\mdot F_2\mdot \sc p \sc p\mdot F_1\mdot F_2\mdot \sc p (a\mdot a)^3+ p_1\mdot \sc p a\mdot F_2\mdot p_1 p_2\mdot F_1\mdot \sc p \sc p\mdot F_1\mdot F_2\mdot \sc p (a\mdot a)^3\Big)\nn\\
   &+\frac{p_1\mdot \sc p \left(m^2 a\mdot p_1 \left(936 a\mdot p_1-883 a\mdot p_2\right)-483 a\mdot a \left(p_1\mdot \sc p\right){}^2\right) p_1\mdot F_2\mdot F_1\mdot \sc p \sc p\mdot F_1\mdot F_2\mdot \sc p (a\mdot a)^3}{11340 m^4}\nn\\
   &+\frac{p_1\mdot \sc p \left(a\mdot p_2 \left(883 a\mdot p_1-936 a\mdot p_2\right) m^2+483 a\mdot a \left(p_1\mdot \sc p\right){}^2\right) p_2\mdot F_1\mdot F_2\mdot \sc p \sc p\mdot F_1\mdot F_2\mdot \sc p (a\mdot a)^3}{11340 m^4}\nn\\
   &+\frac{\left(p_1\mdot \sc p\right){}^2}{5670 m^4} \Big(-13 m^2 \left(132 \left(a\mdot p_1\right){}^2-223 a\mdot p_2 a\mdot p_1+132 \left(a\mdot p_2\right){}^2\right)\nn\\
   &+1987 a\mdot a \left(p_1\mdot \sc p\right){}^2\Big) a\mdot F_1\mdot F_2\mdot a \sc p\mdot F_1\mdot F_2\mdot \sc p (a\mdot a)^2\nn\\
   &+\frac{\left(p_1\mdot \sc p\right){}^2 \left(2 m^2 \left(255 \left(a\mdot p_1\right){}^2-443 a\mdot p_2 a\mdot p_1+255 \left(a\mdot p_2\right){}^2\right)-487 a\mdot a \left(p_1\mdot \sc p\right){}^2\right) \left(a\mdot F_1\mdot F_2\mdot a\right){}^2 a\mdot a}{5670 m^2}\nn\\
   &+\frac{a\mdot p_1 p_1\mdot \sc p \left(3 m^2 a\mdot p_1 \left(33 a\mdot p_1-73 a\mdot p_2\right)-859 a\mdot a \left(p_1\mdot \sc p\right){}^2\right) a\mdot F_1\mdot F_2\mdot a a\mdot F_1\mdot F_2\mdot \sc p a\mdot a}{1890 m^2}\nn\\
   &+\frac{a\mdot p_2 p_1\mdot \sc p \left(3 a\mdot p_2 \left(73 a\mdot p_1-33 a\mdot p_2\right) m^2+859 a\mdot a \left(p_1\mdot \sc p\right){}^2\right) a\mdot F_1\mdot F_2\mdot a a\mdot F_2\mdot F_1\mdot \sc p a\mdot a}{1890 m^2}\nn\\
   &+\frac{(a\mdot a)^2 p_1\mdot \sc p }{5670 m^4}\Big(3 \left(-72 \left(a\mdot p_1\right){}^3+136 a\mdot p_2 \left(a\mdot p_1\right){}^2+83 \left(a\mdot p_2\right){}^2 a\mdot p_1-27 \left(a\mdot p_2\right){}^3\right) m^2\nn\\
   &+a\mdot a \left(2288 a\mdot p_1+289 a\mdot p_2\right) \left(p_1\mdot \sc p\right){}^2\Big) a\mdot F_1\mdot F_2\mdot \sc p \sc p\mdot F_1\mdot F_2\mdot \sc p\nn\\
   &+\frac{(a\mdot a)^2 p_1\mdot \sc p}{5670 m^4} \Big(3 m^2 \left(27 \left(a\mdot p_1\right){}^3-83 a\mdot p_2 \left(a\mdot p_1\right){}^2-136 \left(a\mdot p_2\right){}^2 a\mdot p_1+72 \left(a\mdot p_2\right){}^3\right)\nn\\
   &-a\mdot a \left(289 a\mdot p_1+2288 a\mdot p_2\right) \left(p_1\mdot \sc p\right){}^2\Big) a\mdot F_2\mdot F_1\mdot \sc p \sc p\mdot F_1\mdot F_2\mdot \sc p
\end{align}

\section{UV behaviours}\label{app:UV}
Here we discuss the UV behaviour of the one-loop integral~\eqref{def:M1loop}. Before expanding in spin, the potential divergences can be avoided, when we shift the spin vector to the imaginary axis $a_j \rightarrow i\tilde a_j$. We first illustrate this process with a simple example. Consider the double-copy contribution to the Compton amplitude eq.~\eqref{amplitude structure}, of which the spin-dependence is given in eq.~\eqref{eq:NaDC}. Naively, the worst UV behaviour would have come from the term with $G_1(x_1)G_2(x_2)$, yielding the highest UV scaling at one loop as follows,
\begin{align}
\int {d^D \ell_1\over \pi^{D/2}} {\delta(m_2 v_2\mdot \ell_1) \over \ell_1^2 \ell_2^2} G_1(\ell_1\mdot a_1) G_1 (\ell_2\mdot a_1) \cosh(\ell_1\mdot a_2) \cosh(\ell_2\mdot a_2)\ell_1^\mu \ell_1^\nu,
\end{align}
where we have implicitly $\ell_2=q-\ell_1$ as always. Performing the shift $a_j^\mu= i\tilde a_j^\mu$, the exponential of $\ell_1\mdot a_j$ becomes finite. That is, we have the following scaling as $\ell_1 \rightarrow \Lambda \ell_1$ and $\Lambda\rightarrow\infty$,
\begin{align}
\cosh(\ell_1\mdot a_j)\sim\mathcal O(\Lambda^0),\quad\quad G_1(\ell_1\mdot a_j)\sim\mathcal O(\Lambda^{-1}),\quad\quad G_2(\ell_1\mdot a_j, \ell_2\mdot a_j)\sim \mathcal O(\Lambda^{-1}).
\end{align}
This way, we see that the one-loop integral above is well regulated.

We note that the particular derivatives of the entire functions appearing in eq.~\eqref{amplitude structure} are more suppressed, namely
\begin{align}
  &   (\partial_{a_1\mdot \ell _1}-\partial_{a_1\mdot \ell _2})G_2(a_1\mdot \ell _1,a_1\mdot \ell _2)\sim \mathcal O(\Lambda^{-2}), \\
  & (\partial_{a_1\mdot \ell _1}-\partial_{a_1\mdot \ell _2})\Big(G_1(a_1\mdot \ell _1)G_1(a_1\mdot \ell _2)\Big)\sim \mathcal O(\Lambda^{-2}).
\end{align}

Since all the entire functions involved in~eq.~\eqref{amplitude structure} are all finite under the shift of the spin vector, the UV scaling boils down to the remaining factors given in section~\ref{sec:Compton}, which can be straightforwardly examined. We find that all the terms in~eq.~\eqref{def:M1loop} are indeed well-defined under the shift in a similar fashion, which allows us to evaluate it. In the end, we can analytically continue back to the original real values of the spin vectors.

\section{Decomposition of tensor integrals}\label{app:TenDec}
Here we derive the closed-form expression for the decompositions of the tensor integrals~(\ref{eq:explicitMc5Mc6}). The variables given in eq.~\eqref{eq:tenVar} satisfy the orthogonality relations below,
\begin{align}
   q\mdot \theta_i = \theta_1\mdot\theta_2 = 0, \quad 
   v_i \mdot \theta_j = \delta_{ij},\quad
   q_\mu \Pi^{\mu\nu} = \theta_{i\mu} \Pi^{\mu\nu} = v_{i\mu}\Pi^{\mu\nu} = 0.
\end{align}
This yields the orthogonality of the tensor basis constructed from these building blocks.

We denote the (tensor) integrals as 
\begin{align}
   \mathcal I_{\alpha_1,\alpha_2,\alpha_3,\alpha_4}[f(\ell_1)] \equiv \int {d^D\ell_1 \over \pi^{D/2}} {\delta^{(\alpha_4-1)}(\ell_1\mdot v_2) f(\ell_1) \over \ell_1^{2\alpha_1} (q-\ell_1)^{2\alpha_2} (\ell_1\mdot v_1)^{\alpha_3}}.
\end{align}
We consider the special case where we only have non-vanishing integrals with $\alpha_1 = \alpha_2 = \alpha_4 =1$, since the integrand in eq.~\eqref{def:M1loop} can not have higher propagator powers in $\alpha_1$ or $\alpha_2$ and ultra-local terms (negative values of $\alpha_1$ or $\alpha_2$) are excluded. The scalar integral family is simply $\mathcal I_{1,1,\alpha_3,1}[1]$. The two types of integrals in (\ref{eq:intTenStr}) are $\mathcal I_{1,1,\alpha_3,1}[(\ell_1\mdot a_1)^M (\ell_1\mdot a_2)^{N-M}]$ and $\mathcal I_{1,1,\alpha_3,1}[(\ell_1\mdot a_1)^M (\ell_1\mdot a_2)^{N-M} \ell^\mu]$. 

We consider the tensor numerator with free indices first. The most general ansatz for the decomposition of $\mathcal I_{1,1,\alpha_3,1}[\ell_1^{\mu_1} \cdots \ell_1^{\mu_N}]$ reads
\begin{align}\label{eq:DecAnsatz}
    \mathcal I_{1,1,\alpha_3,1}[\ell_1^{\mu_1} \cdots \ell_1^{\mu_N}] = \sum_{\substack{n_1+n_2+2n_3=N, \\ n_i\geqslant 0, n_i\in\mathbb Z}}& \mathbb C_{n_1 n_2 n_3}\, \text{sym}\left[ \theta_1^{\mu_1} \cdots \theta_1^{\mu_{n_1}} q^{\mu_{n_1+1}} \cdots q^{\mu_{n_1+n_2}} \right. \nn\\
&   \quad\quad \left. \Pi^{\mu_{n_1+n_2+1} \mu_{n_1+n_2+2}} \cdots \Pi^{\mu_{N-1} \mu_N} \right],
\end{align}
where $\text{sym}[\cdots]$ denotes the sum of distinct tensor structures obtained from index permutations. Generally speaking, the coefficients $\mathbb C_{n_1 n_2 n_3}$ can be obtained by contracting both sides of the ansatz with all possible rank-$n$ tensor structures built from $\{v_i^\mu, q_i^\mu, \Pi^{\mu\nu} \}$ and solving the linear equations. 
Thanks to the orthogonality of the tensor structures, these equations are already diagnolized and we can simply read off the coefficients $\mathbb C_{n_1 n_2 n_3}$ as follows,
\begin{align}
    \mathbb C_{n_1 n_2 n_3} = {1\over N_{n_3}} \mathcal I_{1,1,\alpha_3,1}\left[(\ell_1\mdot v_1)^{n_1} (\ell_1\mdot q)^{n_2} (\ell_1\mdot\Pi\mdot\ell_1)^{n_3} \right],
\end{align}
where we have $N_{n_3} = (D-3) (D-1) \cdots (D-3+2(n_3-1))$ for $n_3 >0$ and $N_0 =1$. The remaining integral can be rewritten in terms of scalar integrals,
\begin{align}
   \mathcal I_{1,1,\alpha_3, 1}\left[(\ell_1\mdot v_1)^{n_1} (\ell_1\mdot q)^{n_2} (\ell_1\mdot\Pi\mdot\ell_1)^{n_3} \right] =  \sum_{k=0}^{n_3} \left(\begin{array}{c}
         n_3  \\
          k
    \end{array}\right) {1\over (\gamma^2 -1)^k} \left({q^2 \over 2}\right)^{n_2+n_3-k} \mathcal I_{1,1,\alpha_3-n_1-2k,1},
\end{align}
where we have repeatedly used $\ell_1\mdot q= {1\over 2}\left( \ell_1^2 +q^2- (q-\ell_1)^2\right)$ and that scaleless integrals are vanishing.

The contraction between the symmetrized tensor structures in (\ref{eq:DecAnsatz}) and the spin vectors can be worked out. We have
\begin{align}
\text{Cont}(N,M,n_1,n_2)&\equiv \text{sym} [\, \overbrace{\theta_1 \cdots \theta_1}^{n_1}\, \overbrace{q\cdots q}^{n_2}\, \overbrace{\Pi\cdots \Pi}^{n_3}\, ] \underbrace{a_1\cdots a_1}_{M} \underbrace{a_2\cdots a_2}_{N-M} \nn \\
&=  \sum_{\text{cond.}} C_{N,M,n_1,n_2,m_1,m_2,m_3}~ (a_1\mdot\theta_1)^{m_1} (a_2\mdot\theta_1)^{n_1-m_1} (a_1\mdot q)^{m_2} (a_2\mdot q)^{n_2-m_2} \nn\\
& \quad\quad\quad\quad \quad (a_1\mdot\Pi \mdot a_1)^{m_3} (a_1\mdot\Pi\mdot a_2)^{m_4} (a_2\mdot\Pi\mdot a_2)^{n_3-m_3-m_4},
\end{align}
where the sums are taken over the solutions to the conditions below
\begin{align}
    m_1+m_2+2m_3+m_4 =M,\quad
    0\leqslant m_i \leqslant n_i, \quad 0\leqslant m_4 \leqslant n_3, 
    \quad m_i\in\mathbb Z.
\end{align}
The coefficients read
\begin{align}
    C_{N,M,n_1,n_2,m_1,m_2,m_3} = { M! \over m_1! m_2! m_3! m_4!} {(N-M)! \over (n_1-m_1)! (n_2-m_2)! (n_3-m_3)!} {1\over 2^{n_3-m_4}} .
\end{align}
Hence we have arrived at the decomposition of $\mathcal I_{1,1,\alpha_3,1}[(\ell_1\mdot a_1)^M (\ell_1\mdot a_2)^{N-M}]$. 

The remaining integral $\mathcal I_{1,1,\alpha_3,1}[(\ell_1\mdot a_1)^M (\ell_1\mdot a_2)^{N-M} \ell_1^\mu]$ is nothing but the above with one free index and can be dealt with similarly. We first rewrite the symmeterized sum of the tensor structures as follows,
\begin{align}
    \text{sym} [\, \overbrace{\theta_1 \cdots \theta_1}^{n_1}\, \overbrace{q\cdots q}^{n_2}\, \overbrace{\Pi\cdots \Pi}^{n_3}\, ] = &\quad \theta_1^{\mu_N} \text{sym} [\, \overbrace{\theta_1 \cdots \theta_1}^{n_1-1}\, \overbrace{q\cdots q}^{n_2}\, \overbrace{\Pi\cdots \Pi}^{n_3}\, ] \nn \\ 
    &+ q^{\mu_N}  \text{sym} [\, \overbrace{\theta_1 \cdots \theta_1}^{n_1}\, \overbrace{q\cdots q}^{n_2-1}\, \overbrace{\Pi\cdots \Pi}^{n_3}\, ] \nn \\
   & +\sum_{i=1}^{N-1} \Pi^{\mu_i \mu_N}  \text{sym} [\, \overbrace{\theta_1 \cdots \theta_1}^{n_1}\, \overbrace{q\cdots q}^{n_2}\, \overbrace{\Pi\cdots \Pi}^{n_3-1}\, ].
\end{align}
Contracting with $a_1^{\mu_1}\cdots a_1^{\mu_M} a_2^{\mu_{M+1}}\cdots a_2^{\mu_{N-M-1}}$, we have
\begin{align}
&  \quad  \text{sym} [\, \overbrace{\theta_1 \cdots \theta_1}^{n_1}\, \overbrace{q\cdots q}^{n_2}\, \overbrace{\Pi\cdots \Pi}^{n_3}\, ] \underbrace{a_1\cdots a_1}_{M} \underbrace{a_2\cdots a_2}_{N-1-M} \nn \\
=&\quad  \theta^{\mu_N}_1 \text{Cont}(N-1,M,n_1-1,n_2) + q^{\mu_N} \text{Cont}(N-1,M,n_1,n_2-1) \\
 & + M (a_1\mdot\Pi)^{\mu_N} \text{Cont}(N-2,M-1,n_1,n_2) + (N-1-M)(a_2\mdot\Pi)^{\mu_N} \text{Cont}(N-2,M,n_1,n_2).\nn
\end{align}

\bibliographystyle{JHEP}
\bibliography{spin2pm}

\providecommand{\href}[2]{#2}\begingroup\raggedright\begin{thebibliography}{100}

\bibitem{LIGOScientific:2016dsl}
{\scshape LIGO Scientific, Virgo} collaboration, \emph{{Binary Black Hole
  Mergers in the first Advanced LIGO Observing Run}},
  \href{https://doi.org/10.1103/PhysRevX.6.041015}{\emph{Phys. Rev. X}
  {\bfseries 6} (2016) 041015}
  [\href{https://arxiv.org/abs/1606.04856}{{\ttfamily 1606.04856}}].

\bibitem{LIGOScientific:2016aoc}
{\scshape LIGO Scientific, Virgo} collaboration, \emph{{Observation of
  Gravitational Waves from a Binary Black Hole Merger}},
  \href{https://doi.org/10.1103/PhysRevLett.116.061102}{\emph{Phys. Rev. Lett.}
  {\bfseries 116} (2016) 061102}
  [\href{https://arxiv.org/abs/1602.03837}{{\ttfamily 1602.03837}}].

\bibitem{LIGOScientific:2016sjg}
{\scshape LIGO Scientific, Virgo} collaboration, \emph{{GW151226: Observation
  of Gravitational Waves from a 22-Solar-Mass Binary Black Hole Coalescence}},
  \href{https://doi.org/10.1103/PhysRevLett.116.241103}{\emph{Phys. Rev. Lett.}
  {\bfseries 116} (2016) 241103}
  [\href{https://arxiv.org/abs/1606.04855}{{\ttfamily 1606.04855}}].

\bibitem{LIGOScientific:2017bnn}
{\scshape LIGO Scientific, VIRGO} collaboration, \emph{{GW170104: Observation
  of a 50-Solar-Mass Binary Black Hole Coalescence at Redshift 0.2}},
  \href{https://doi.org/10.1103/PhysRevLett.118.221101}{\emph{Phys. Rev. Lett.}
  {\bfseries 118} (2017) 221101}
  [\href{https://arxiv.org/abs/1706.01812}{{\ttfamily 1706.01812}}].

\bibitem{LIGOScientific:2017vwq}
{\scshape LIGO Scientific, Virgo} collaboration, \emph{{GW170817: Observation
  of Gravitational Waves from a Binary Neutron Star Inspiral}},
  \href{https://doi.org/10.1103/PhysRevLett.119.161101}{\emph{Phys. Rev. Lett.}
  {\bfseries 119} (2017) 161101}
  [\href{https://arxiv.org/abs/1710.05832}{{\ttfamily 1710.05832}}].

\bibitem{Pretorius:2005gq}
F.~Pretorius, \emph{{Evolution of binary black hole spacetimes}},
  \href{https://doi.org/10.1103/PhysRevLett.95.121101}{\emph{Phys. Rev. Lett.}
  {\bfseries 95} (2005) 121101}
  [\href{https://arxiv.org/abs/gr-qc/0507014}{{\ttfamily gr-qc/0507014}}].

\bibitem{Campanelli:2005dd}
M.~Campanelli, C.O.~Lousto, P.~Marronetti and Y.~Zlochower, \emph{{Accurate
  evolutions of orbiting black-hole binaries without excision}},
  \href{https://doi.org/10.1103/PhysRevLett.96.111101}{\emph{Phys. Rev. Lett.}
  {\bfseries 96} (2006) 111101}
  [\href{https://arxiv.org/abs/gr-qc/0511048}{{\ttfamily gr-qc/0511048}}].

\bibitem{Baker:2005vv}
J.G.~Baker, J.~Centrella, D.-I.~Choi, M.~Koppitz and J.~van Meter,
  \emph{{Gravitational wave extraction from an inspiraling configuration of
  merging black holes}},
  \href{https://doi.org/10.1103/PhysRevLett.96.111102}{\emph{Phys. Rev. Lett.}
  {\bfseries 96} (2006) 111102}
  [\href{https://arxiv.org/abs/gr-qc/0511103}{{\ttfamily gr-qc/0511103}}].

\bibitem{Buonanno:1998gg}
A.~Buonanno and T.~Damour, \emph{{Effective one-body approach to general
  relativistic two-body dynamics}},
  \href{https://doi.org/10.1103/PhysRevD.59.084006}{\emph{Phys. Rev.}
  {\bfseries D59} (1999) 084006}
  [\href{https://arxiv.org/abs/gr-qc/9811091}{{\ttfamily gr-qc/9811091}}].

\bibitem{Goldberger:2004jt}
W.D.~Goldberger and I.Z.~Rothstein, \emph{{An Effective field theory of gravity
  for extended objects}},
  \href{https://doi.org/10.1103/PhysRevD.73.104029}{\emph{Phys. Rev. D}
  {\bfseries 73} (2006) 104029}
  [\href{https://arxiv.org/abs/hep-th/0409156}{{\ttfamily hep-th/0409156}}].

\bibitem{Kol:2007rx}
B.~Kol and M.~Smolkin, \emph{{Classical Effective Field Theory and Caged Black
  Holes}}, \href{https://doi.org/10.1103/PhysRevD.77.064033}{\emph{Phys. Rev.
  D} {\bfseries 77} (2008) 064033}
  [\href{https://arxiv.org/abs/0712.2822}{{\ttfamily 0712.2822}}].

\bibitem{Gilmore:2008gq}
J.B.~Gilmore and A.~Ross, \emph{{Effective field theory calculation of second
  post-Newtonian binary dynamics}},
  \href{https://doi.org/10.1103/PhysRevD.78.124021}{\emph{Phys. Rev. D}
  {\bfseries 78} (2008) 124021}
  [\href{https://arxiv.org/abs/0810.1328}{{\ttfamily 0810.1328}}].

\bibitem{Foffa:2011ub}
S.~Foffa and R.~Sturani, \emph{{Effective field theory calculation of
  conservative binary dynamics at third post-Newtonian order}},
  \href{https://doi.org/10.1103/PhysRevD.84.044031}{\emph{Phys. Rev. D}
  {\bfseries 84} (2011) 044031}
  [\href{https://arxiv.org/abs/1104.1122}{{\ttfamily 1104.1122}}].

\bibitem{Foffa:2016rgu}
S.~Foffa, P.~Mastrolia, R.~Sturani and C.~Sturm, \emph{{Effective field theory
  approach to the gravitational two-body dynamics, at fourth post-Newtonian
  order and quintic in the Newton constant}},
  \href{https://doi.org/10.1103/PhysRevD.95.104009}{\emph{Phys. Rev. D}
  {\bfseries 95} (2017) 104009}
  [\href{https://arxiv.org/abs/1612.00482}{{\ttfamily 1612.00482}}].

\bibitem{Porto:2017dgs}
R.A.~Porto and I.Z.~Rothstein, \emph{{Apparent ambiguities in the
  post-Newtonian expansion for binary systems}},
  \href{https://doi.org/10.1103/PhysRevD.96.024062}{\emph{Phys. Rev. D}
  {\bfseries 96} (2017) 024062}
  [\href{https://arxiv.org/abs/1703.06433}{{\ttfamily 1703.06433}}].

\bibitem{Blumlein:2019zku}
J.~Bl\"umlein, A.~Maier and P.~Marquard, \emph{{Five-Loop Static Contribution
  to the Gravitational Interaction Potential of Two Point Masses}},
  \href{https://doi.org/10.1016/j.physletb.2019.135100}{\emph{Phys. Lett. B}
  {\bfseries 800} (2020) 135100}
  [\href{https://arxiv.org/abs/1902.11180}{{\ttfamily 1902.11180}}].

\bibitem{Foffa:2019rdf}
S.~Foffa and R.~Sturani, \emph{{Conservative dynamics of binary systems to
  fourth Post-Newtonian order in the EFT approach I: Regularized Lagrangian}},
  \href{https://doi.org/10.1103/PhysRevD.100.024047}{\emph{Phys. Rev. D}
  {\bfseries 100} (2019) 024047}
  [\href{https://arxiv.org/abs/1903.05113}{{\ttfamily 1903.05113}}].

\bibitem{Foffa:2019yfl}
S.~Foffa, R.A.~Porto, I.~Rothstein and R.~Sturani, \emph{{Conservative dynamics
  of binary systems to fourth Post-Newtonian order in the EFT approach II:
  Renormalized Lagrangian}},
  \href{https://doi.org/10.1103/PhysRevD.100.024048}{\emph{Phys. Rev. D}
  {\bfseries 100} (2019) 024048}
  [\href{https://arxiv.org/abs/1903.05118}{{\ttfamily 1903.05118}}].

\bibitem{Blumlein:2020pog}
J.~Bl\"umlein, A.~Maier, P.~Marquard and G.~Sch\"afer, \emph{{Fourth
  post-Newtonian Hamiltonian dynamics of two-body systems from an effective
  field theory approach}},
  \href{https://doi.org/10.1016/j.nuclphysb.2020.115041}{\emph{Nucl. Phys. B}
  {\bfseries 955} (2020) 115041}
  [\href{https://arxiv.org/abs/2003.01692}{{\ttfamily 2003.01692}}].

\bibitem{Blumlein:2020znm}
J.~Bluemlein, A.~Maier, P.~Marquard and G.~Schaefer, \emph{{Testing binary
  dynamics in gravity at the sixth post-Newtonian level}},
  \href{https://arxiv.org/abs/2003.07145}{{\ttfamily 2003.07145}}.

\bibitem{Bini:2020nsb}
D.~Bini, T.~Damour and A.~Geralico, \emph{{Sixth post-Newtonian local-in-time
  dynamics of binary systems}},
  \href{https://doi.org/10.1103/PhysRevD.102.024061}{\emph{Phys. Rev. D}
  {\bfseries 102} (2020) 024061}
  [\href{https://arxiv.org/abs/2004.05407}{{\ttfamily 2004.05407}}].

\bibitem{Bini:2020hmy}
D.~Bini, T.~Damour and A.~Geralico, \emph{{Sixth post-Newtonian
  nonlocal-in-time dynamics of binary systems}},
  \href{https://doi.org/10.1103/PhysRevD.102.084047}{\emph{Phys. Rev. D}
  {\bfseries 102} (2020) 084047}
  [\href{https://arxiv.org/abs/2007.11239}{{\ttfamily 2007.11239}}].

\bibitem{Blumlein:2021txe}
J.~Bl\"umlein, A.~Maier, P.~Marquard and G.~Sch\"afer, \emph{{The fifth-order
  post-Newtonian Hamiltonian dynamics of two-body systems from an effective
  field theory approach}},
  \href{https://doi.org/10.1016/j.nuclphysb.2022.115900}{\emph{Nucl. Phys. B}
  {\bfseries 983} (2022) 115900}
  [\href{https://arxiv.org/abs/2110.13822}{{\ttfamily 2110.13822}}].

\bibitem{Blumlein:2020pyo}
J.~Bl\"umlein, A.~Maier, P.~Marquard and G.~Sch\"afer, \emph{{The fifth-order
  post-Newtonian Hamiltonian dynamics of two-body systems from an effective
  field theory approach: potential contributions}},
  \href{https://arxiv.org/abs/2010.13672}{{\ttfamily 2010.13672}}.

\bibitem{Foffa:2020nqe}
S.~Foffa, R.~Sturani and W.J.~Torres~Bobadilla, \emph{{Efficient resummation of
  high post-Newtonian contributions to the binding energy}},
  \href{https://doi.org/10.1007/JHEP02(2021)165}{\emph{JHEP} {\bfseries 02}
  (2021) 165} [\href{https://arxiv.org/abs/2010.13730}{{\ttfamily
  2010.13730}}].

\bibitem{Blumlein:2021txj}
J.~Bl\"umlein, A.~Maier, P.~Marquard and G.~Sch\"afer, \emph{{The 6th
  post-Newtonian potential terms at $O(G_N^4)$}},
  \href{https://doi.org/10.1016/j.physletb.2021.136260}{\emph{Phys. Lett. B}
  {\bfseries 816} (2021) 136260}
  [\href{https://arxiv.org/abs/2101.08630}{{\ttfamily 2101.08630}}].

\bibitem{Kim:2021rfj}
J.-W.~Kim, M.~Levi and Z.~Yin, \emph{{Quadratic-in-spin interactions at fifth
  post-Newtonian order probe new physics}},
  \href{https://doi.org/10.1016/j.physletb.2022.137410}{\emph{Phys. Lett. B}
  {\bfseries 834} (2022) 137410}
  [\href{https://arxiv.org/abs/2112.01509}{{\ttfamily 2112.01509}}].

\bibitem{Cho:2022syn}
G.~Cho, R.A.~Porto and Z.~Yang, \emph{{Gravitational radiation from
  inspiralling compact objects: Spin effects to fourth Post-Newtonian order}},
  \href{https://arxiv.org/abs/2201.05138}{{\ttfamily 2201.05138}}.

\bibitem{Bjerrum-Bohr:2018xdl}
N.E.J.~Bjerrum-Bohr, P.H.~Damgaard, G.~Festuccia, L.~Plante and P.~Vanhove,
  \emph{{General Relativity from Scattering Amplitudes}},
  \href{https://doi.org/10.1103/PhysRevLett.121.171601}{\emph{Phys. Rev. Lett.}
  {\bfseries 121} (2018) 171601}
  [\href{https://arxiv.org/abs/1806.04920}{{\ttfamily 1806.04920}}].

\bibitem{Cheung:2018wkq}
C.~Cheung, I.Z.~Rothstein and M.P.~Solon, \emph{{From Scattering Amplitudes to
  Classical Potentials in the Post-Minkowskian Expansion}},
  \href{https://doi.org/10.1103/PhysRevLett.121.251101}{\emph{Phys. Rev. Lett.}
  {\bfseries 121} (2018) 251101}
  [\href{https://arxiv.org/abs/1808.02489}{{\ttfamily 1808.02489}}].

\bibitem{Bern:2019nnu}
Z.~Bern, C.~Cheung, R.~Roiban, C.-H.~Shen, M.P.~Solon and M.~Zeng,
  \emph{{Scattering Amplitudes and the Conservative Hamiltonian for Binary
  Systems at Third Post-Minkowskian Order}},
  \href{https://doi.org/10.1103/PhysRevLett.122.201603}{\emph{Phys. Rev. Lett.}
  {\bfseries 122} (2019) 201603}
  [\href{https://arxiv.org/abs/1901.04424}{{\ttfamily 1901.04424}}].

\bibitem{Bern:2019crd}
Z.~Bern, C.~Cheung, R.~Roiban, C.-H.~Shen, M.P.~Solon and M.~Zeng, \emph{{Black
  Hole Binary Dynamics from the Double Copy and Effective Theory}},
  \href{https://doi.org/10.1007/JHEP10(2019)206}{\emph{JHEP} {\bfseries 10}
  (2019) 206} [\href{https://arxiv.org/abs/1908.01493}{{\ttfamily
  1908.01493}}].

\bibitem{Neill:2013wsa}
D.~Neill and I.Z.~Rothstein, \emph{{Classical Space-Times from the S Matrix}},
  \href{https://doi.org/10.1016/j.nuclphysb.2013.09.007}{\emph{Nucl. Phys.}
  {\bfseries B877} (2013) 177}
  [\href{https://arxiv.org/abs/1304.7263}{{\ttfamily 1304.7263}}].

\bibitem{Cristofoli:2021vyo}
A.~Cristofoli, R.~Gonzo, D.A.~Kosower and D.~O'Connell, \emph{{Waveforms from
  Amplitudes}},  \href{https://arxiv.org/abs/2107.10193}{{\ttfamily
  2107.10193}}.

\bibitem{Bern:2020buy}
Z.~Bern, A.~Luna, R.~Roiban, C.-H.~Shen and M.~Zeng, \emph{{Spinning black hole
  binary dynamics, scattering amplitudes, and effective field theory}},
  \href{https://doi.org/10.1103/PhysRevD.104.065014}{\emph{Phys. Rev. D}
  {\bfseries 104} (2021) 065014}
  [\href{https://arxiv.org/abs/2005.03071}{{\ttfamily 2005.03071}}].

\bibitem{Bern:2021dqo}
Z.~Bern, J.~Parra-Martinez, R.~Roiban, M.S.~Ruf, C.-H.~Shen, M.P.~Solon et~al.,
  \emph{{Scattering Amplitudes and Conservative Binary Dynamics at ${\cal
  O}(G^4)$}}, \href{https://doi.org/10.1103/PhysRevLett.126.171601}{\emph{Phys.
  Rev. Lett.} {\bfseries 126} (2021) 171601}
  [\href{https://arxiv.org/abs/2101.07254}{{\ttfamily 2101.07254}}].

\bibitem{Bjerrum-Bohr:2002gqz}
N.E.J.~Bjerrum-Bohr, J.F.~Donoghue and B.R.~Holstein, \emph{{Quantum
  gravitational corrections to the nonrelativistic scattering potential of two
  masses}}, \href{https://doi.org/10.1103/PhysRevD.71.069903}{\emph{Phys. Rev.
  D} {\bfseries 67} (2003) 084033}
  [\href{https://arxiv.org/abs/hep-th/0211072}{{\ttfamily hep-th/0211072}}].

\bibitem{Kosower:2018adc}
D.A.~Kosower, B.~Maybee and D.~O'Connell, \emph{{Amplitudes, Observables, and
  Classical Scattering}},
  \href{https://doi.org/10.1007/JHEP02(2019)137}{\emph{JHEP} {\bfseries 02}
  (2019) 137} [\href{https://arxiv.org/abs/1811.10950}{{\ttfamily
  1811.10950}}].

\bibitem{Maybee:2019jus}
B.~Maybee, D.~O'Connell and J.~Vines, \emph{{Observables and amplitudes for
  spinning particles and black holes}},
  \href{https://doi.org/10.1007/JHEP12(2019)156}{\emph{JHEP} {\bfseries 12}
  (2019) 156} [\href{https://arxiv.org/abs/1906.09260}{{\ttfamily
  1906.09260}}].

\bibitem{Brandhuber:2021kpo}
A.~Brandhuber, G.~Chen, G.~Travaglini and C.~Wen, \emph{{A new gauge-invariant
  double copy for heavy-mass effective theory}},
  \href{https://doi.org/10.1007/JHEP07(2021)047}{\emph{JHEP} {\bfseries 07}
  (2021) 047} [\href{https://arxiv.org/abs/2104.11206}{{\ttfamily
  2104.11206}}].

\bibitem{Brandhuber:2021eyq}
A.~Brandhuber, G.~Chen, G.~Travaglini and C.~Wen, \emph{{Classical
  gravitational scattering from a gauge-invariant double copy}},
  \href{https://doi.org/10.1007/JHEP10(2021)118}{\emph{JHEP} {\bfseries 10}
  (2021) 118} [\href{https://arxiv.org/abs/2108.04216}{{\ttfamily
  2108.04216}}].

\bibitem{Mogull:2020sak}
G.~Mogull, J.~Plefka and J.~Steinhoff, \emph{{Classical black hole scattering
  from a worldline quantum field theory}},
  \href{https://doi.org/10.1007/JHEP02(2021)048}{\emph{JHEP} {\bfseries 02}
  (2021) 048} [\href{https://arxiv.org/abs/2010.02865}{{\ttfamily
  2010.02865}}].

\bibitem{Jakobsen:2021smu}
G.U.~Jakobsen, G.~Mogull, J.~Plefka and J.~Steinhoff, \emph{{Classical
  Gravitational Bremsstrahlung from a Worldline Quantum Field Theory}},
  \href{https://doi.org/10.1103/PhysRevLett.126.201103}{\emph{Phys. Rev. Lett.}
  {\bfseries 126} (2021) 201103}
  [\href{https://arxiv.org/abs/2101.12688}{{\ttfamily 2101.12688}}].

\bibitem{Jakobsen:2022fcj}
G.U.~Jakobsen and G.~Mogull, \emph{{Conservative and Radiative Dynamics of
  Spinning Bodies at Third Post-Minkowskian Order Using Worldline Quantum Field
  Theory}}, \href{https://doi.org/10.1103/PhysRevLett.128.141102}{\emph{Phys.
  Rev. Lett.} {\bfseries 128} (2022) 141102}
  [\href{https://arxiv.org/abs/2201.07778}{{\ttfamily 2201.07778}}].

\bibitem{Wang:2022ntx}
T.~Wang, \emph{{Binary dynamics from worldline QFT for scalar QED}},
  \href{https://doi.org/10.1103/PhysRevD.107.085011}{\emph{Phys. Rev. D}
  {\bfseries 107} (2023) 085011}
  [\href{https://arxiv.org/abs/2205.15753}{{\ttfamily 2205.15753}}].

\bibitem{Parra-Martinez:2020dzs}
J.~Parra-Martinez, M.S.~Ruf and M.~Zeng, \emph{{Extremal black hole scattering
  at $\mathcal{O}(G^3)$: graviton dominance, eikonal exponentiation, and
  differential equations}},
  \href{https://doi.org/10.1007/JHEP11(2020)023}{\emph{JHEP} {\bfseries 11}
  (2020) 023} [\href{https://arxiv.org/abs/2005.04236}{{\ttfamily
  2005.04236}}].

\bibitem{DiVecchia:2021bdo}
P.~Di~Vecchia, C.~Heissenberg, R.~Russo and G.~Veneziano, \emph{{The eikonal
  approach to gravitational scattering and radiation at $ \mathcal{O}
  $(G$^{3}$)}}, \href{https://doi.org/10.1007/JHEP07(2021)169}{\emph{JHEP}
  {\bfseries 07} (2021) 169}
  [\href{https://arxiv.org/abs/2104.03256}{{\ttfamily 2104.03256}}].

\bibitem{Heissenberg:2021tzo}
C.~Heissenberg, \emph{{Infrared divergences and the eikonal exponentiation}},
  \href{https://doi.org/10.1103/PhysRevD.104.046016}{\emph{Phys. Rev. D}
  {\bfseries 104} (2021) 046016}
  [\href{https://arxiv.org/abs/2105.04594}{{\ttfamily 2105.04594}}].

\bibitem{DiVecchia:2022nna}
P.~Di~Vecchia, C.~Heissenberg, R.~Russo and G.~Veneziano, \emph{{The eikonal
  operator at arbitrary velocities I: the soft-radiation limit}},
  \href{https://arxiv.org/abs/2204.02378}{{\ttfamily 2204.02378}}.

\bibitem{Damgaard:2021ipf}
P.H.~Damgaard, L.~Plante and P.~Vanhove, \emph{{On an exponential
  representation of the gravitational S-matrix}},
  \href{https://doi.org/10.1007/JHEP11(2021)213}{\emph{JHEP} {\bfseries 11}
  (2021) 213} [\href{https://arxiv.org/abs/2107.12891}{{\ttfamily
  2107.12891}}].

\bibitem{Kol:2021jjc}
U.~Kol, D.~O'connell and O.~Telem, \emph{{The radial action from probe
  amplitudes to all orders}},
  \href{https://doi.org/10.1007/JHEP03(2022)141}{\emph{JHEP} {\bfseries 03}
  (2022) 141} [\href{https://arxiv.org/abs/2109.12092}{{\ttfamily
  2109.12092}}].

\bibitem{Bjerrum-Bohr:2021wwt}
N.E.J.~Bjerrum-Bohr, L.~Plant\'e and P.~Vanhove, \emph{{Post-Minkowskian radial
  action from soft limits and velocity cuts}},
  \href{https://doi.org/10.1007/JHEP03(2022)071}{\emph{JHEP} {\bfseries 03}
  (2022) 071} [\href{https://arxiv.org/abs/2111.02976}{{\ttfamily
  2111.02976}}].

\bibitem{Bern:2021xze}
Z.~Bern, J.P.~Gatica, E.~Herrmann, A.~Luna and M.~Zeng, \emph{{Scalar QED as a
  toy model for higher-order effects in classical gravitational scattering}},
  \href{https://arxiv.org/abs/2112.12243}{{\ttfamily 2112.12243}}.

\bibitem{Lee:2023vdy}
H.~Lee, S.~Lee and S.~Mazumdar, \emph{{Classical observables from partial wave
  amplitudes}}, \href{https://doi.org/10.1007/JHEP06(2023)096}{\emph{JHEP}
  {\bfseries 06} (2023) 096}
  [\href{https://arxiv.org/abs/2303.07638}{{\ttfamily 2303.07638}}].

\bibitem{Lee:2023zuu}
H.~Lee, K.~Lee and S.~Lee, \emph{{Poincar\'e generators at second
  post-Minkowskian order}},
  \href{https://doi.org/10.1007/JHEP10(2023)044}{\emph{JHEP} {\bfseries 10}
  (2023) 044} [\href{https://arxiv.org/abs/2307.05626}{{\ttfamily
  2307.05626}}].

\bibitem{Dlapa:2021npj}
C.~Dlapa, G.~K\"alin, Z.~Liu and R.A.~Porto, \emph{{Dynamics of binary systems
  to fourth Post-Minkowskian order from the effective field theory approach}},
  \href{https://doi.org/10.1016/j.physletb.2022.137203}{\emph{Phys. Lett. B}
  {\bfseries 831} (2022) 137203}
  [\href{https://arxiv.org/abs/2106.08276}{{\ttfamily 2106.08276}}].

\bibitem{Dlapa:2022lmu}
C.~Dlapa, G.~K\"alin, Z.~Liu, J.~Neef and R.A.~Porto, \emph{{Radiation Reaction
  and Gravitational Waves at Fourth Post-Minkowskian Order}},
  \href{https://arxiv.org/abs/2210.05541}{{\ttfamily 2210.05541}}.

\bibitem{Dlapa:2023hsl}
C.~Dlapa, G.~K\"alin, Z.~Liu and R.A.~Porto, \emph{{Bootstrapping the
  relativistic two-body problem}},
  \href{https://doi.org/10.1007/JHEP08(2023)109}{\emph{JHEP} {\bfseries 08}
  (2023) 109} [\href{https://arxiv.org/abs/2304.01275}{{\ttfamily
  2304.01275}}].

\bibitem{Jakobsen:2023pvx}
G.U.~Jakobsen, G.~Mogull, J.~Plefka and B.~Sauer, \emph{{Tidal effects and
  renormalization at fourth post-Minkowskian order}},
  \href{https://doi.org/10.1103/PhysRevD.109.L041504}{\emph{Phys. Rev. D}
  {\bfseries 109} (2024) L041504}
  [\href{https://arxiv.org/abs/2312.00719}{{\ttfamily 2312.00719}}].

\bibitem{Bern:2021yeh}
Z.~Bern, J.~Parra-Martinez, R.~Roiban, M.S.~Ruf, C.-H.~Shen, M.P.~Solon et~al.,
  \emph{{Scattering Amplitudes, the Tail Effect, and Conservative Binary
  Dynamics at O(G4)}},
  \href{https://doi.org/10.1103/PhysRevLett.128.161103}{\emph{Phys. Rev. Lett.}
  {\bfseries 128} (2022) 161103}
  [\href{https://arxiv.org/abs/2112.10750}{{\ttfamily 2112.10750}}].

\bibitem{Damgaard:2023ttc}
P.H.~Damgaard, E.R.~Hansen, L.~Plant\'e and P.~Vanhove, \emph{{Classical
  Observables from the Exponential Representation of the Gravitational
  S-Matrix}},  \href{https://arxiv.org/abs/2307.04746}{{\ttfamily 2307.04746}}.

\bibitem{Driesse:2024xad}
M.~Driesse, G.U.~Jakobsen, G.~Mogull, J.~Plefka, B.~Sauer and J.~Usovitsch,
  \emph{{Conservative Black Hole Scattering at Fifth Post-Minkowskian and First
  Self-Force Order}},  \href{https://arxiv.org/abs/2403.07781}{{\ttfamily
  2403.07781}}.

\bibitem{Arkani-Hamed:2017jhn}
N.~Arkani-Hamed, T.-C.~Huang and Y.-t.~Huang, \emph{{Scattering amplitudes for
  all masses and spins}},
  \href{https://doi.org/10.1007/JHEP11(2021)070}{\emph{JHEP} {\bfseries 11}
  (2021) 070} [\href{https://arxiv.org/abs/1709.04891}{{\ttfamily
  1709.04891}}].

\bibitem{Zinoviev:2001dt}
Y.M.~Zinoviev, \emph{{On massive high spin particles in AdS}},
  \href{https://arxiv.org/abs/hep-th/0108192}{{\ttfamily hep-th/0108192}}.

\bibitem{Zinoviev:2006im}
Y.M.~Zinoviev, \emph{{On massive spin 2 interactions}},
  \href{https://doi.org/10.1016/j.nuclphysb.2007.02.005}{\emph{Nucl. Phys. B}
  {\bfseries 770} (2007) 83}
  [\href{https://arxiv.org/abs/hep-th/0609170}{{\ttfamily hep-th/0609170}}].

\bibitem{Zinoviev:2009hu}
Y.M.~Zinoviev, \emph{{On massive spin 2 electromagnetic interactions}},
  \href{https://doi.org/10.1016/j.nuclphysb.2009.04.027}{\emph{Nucl. Phys. B}
  {\bfseries 821} (2009) 431}
  [\href{https://arxiv.org/abs/0901.3462}{{\ttfamily 0901.3462}}].

\bibitem{Zinoviev:2008ck}
Y.M.~Zinoviev, \emph{{On spin 3 interacting with gravity}},
  \href{https://doi.org/10.1088/0264-9381/26/3/035022}{\emph{Class. Quant.
  Grav.} {\bfseries 26} (2009) 035022}
  [\href{https://arxiv.org/abs/0805.2226}{{\ttfamily 0805.2226}}].

\bibitem{Ochirov:2022nqz}
A.~Ochirov and E.~Skvortsov, \emph{{Chiral Approach to Massive Higher Spins}},
  \href{https://doi.org/10.1103/PhysRevLett.129.241601}{\emph{Phys. Rev. Lett.}
  {\bfseries 129} (2022) 241601}
  [\href{https://arxiv.org/abs/2207.14597}{{\ttfamily 2207.14597}}].

\bibitem{Cangemi:2022abk}
L.~Cangemi and P.~Pichini, \emph{{Classical limit of higher-spin string
  amplitudes}}, \href{https://doi.org/10.1007/JHEP06(2023)167}{\emph{JHEP}
  {\bfseries 06} (2023) 167}
  [\href{https://arxiv.org/abs/2207.03947}{{\ttfamily 2207.03947}}].

\bibitem{Bern:2008qj}
Z.~Bern, J.J.M.~Carrasco and H.~Johansson, \emph{{New Relations for
  Gauge-Theory Amplitudes}},
  \href{https://doi.org/10.1103/PhysRevD.78.085011}{\emph{Phys. Rev.}
  {\bfseries D78} (2008) 085011}
  [\href{https://arxiv.org/abs/0805.3993}{{\ttfamily 0805.3993}}].

\bibitem{Bern:2010ue}
Z.~Bern, J.J.M.~Carrasco and H.~Johansson, \emph{{Perturbative Quantum Gravity
  as a Double Copy of Gauge Theory}},
  \href{https://doi.org/10.1103/PhysRevLett.105.061602}{\emph{Phys. Rev. Lett.}
  {\bfseries 105} (2010) 061602}
  [\href{https://arxiv.org/abs/1004.0476}{{\ttfamily 1004.0476}}].

\bibitem{Bern:2019prr}
Z.~Bern, J.J.~Carrasco, M.~Chiodaroli, H.~Johansson and R.~Roiban, \emph{{The
  Duality Between Color and Kinematics and its Applications}},
  \href{https://arxiv.org/abs/1909.01358}{{\ttfamily 1909.01358}}.

\bibitem{Bjerrum-Bohr:2023jau}
N.E.J.~Bjerrum-Bohr, G.~Chen and M.~Skowronek, \emph{{Classical spin
  gravitational Compton scattering}},
  \href{https://doi.org/10.1007/JHEP06(2023)170}{\emph{JHEP} {\bfseries 06}
  (2023) 170} [\href{https://arxiv.org/abs/2302.00498}{{\ttfamily
  2302.00498}}].

\bibitem{Bjerrum-Bohr:2023iey}
N.E.J.~Bjerrum-Bohr, G.~Chen and M.~Skowronek, \emph{{Covariant Compton
  Amplitudes in Gravity with Classical Spin}},
  \href{https://doi.org/10.1103/PhysRevLett.132.191603}{\emph{Phys. Rev. Lett.}
  {\bfseries 132} (2024) 191603}
  [\href{https://arxiv.org/abs/2309.11249}{{\ttfamily 2309.11249}}].

\bibitem{Teukolsky:1973ha}
S.A.~Teukolsky, \emph{{Perturbations of a rotating black hole. 1. Fundamental
  equations for gravitational electromagnetic and neutrino field
  perturbations}}, \href{https://doi.org/10.1086/152444}{\emph{Astrophys. J.}
  {\bfseries 185} (1973) 635}.

\bibitem{Press:1973zz}
W.H.~Press and S.A.~Teukolsky, \emph{{Perturbations of a Rotating Black Hole.
  II. Dynamical Stability of the Kerr Metric}},
  \href{https://doi.org/10.1086/152445}{\emph{Astrophys. J.} {\bfseries 185}
  (1973) 649}.

\bibitem{Teukolsky:1974yv}
S.A.~Teukolsky and W.H.~Press, \emph{{Perturbations of a rotating black hole.
  III - Interaction of the hole with gravitational and electromagnet ic
  radiation}}, \href{https://doi.org/10.1086/153180}{\emph{Astrophys. J.}
  {\bfseries 193} (1974) 443}.

\bibitem{Chia:2020yla}
H.S.~Chia, \emph{{Tidal deformation and dissipation of rotating black holes}},
  \href{https://doi.org/10.1103/PhysRevD.104.024013}{\emph{Phys. Rev. D}
  {\bfseries 104} (2021) 024013}
  [\href{https://arxiv.org/abs/2010.07300}{{\ttfamily 2010.07300}}].

\bibitem{Bautista:2021wfy}
Y.F.~Bautista, A.~Guevara, C.~Kavanagh and J.~Vines, \emph{{From Scattering in
  Black Hole Backgrounds to Higher-Spin Amplitudes: Part I}},
  \href{https://arxiv.org/abs/2107.10179}{{\ttfamily 2107.10179}}.

\bibitem{Ivanov:2022qqt}
M.M.~Ivanov and Z.~Zhou, \emph{{Vanishing of Black Hole Tidal Love Numbers from
  Scattering Amplitudes}},
  \href{https://doi.org/10.1103/PhysRevLett.130.091403}{\emph{Phys. Rev. Lett.}
  {\bfseries 130} (2023) 091403}
  [\href{https://arxiv.org/abs/2209.14324}{{\ttfamily 2209.14324}}].

\bibitem{Bautista:2022wjf}
Y.F.~Bautista, A.~Guevara, C.~Kavanagh and J.~Vines, \emph{{Scattering in black
  hole backgrounds and higher-spin amplitudes. Part II}},
  \href{https://doi.org/10.1007/JHEP05(2023)211}{\emph{JHEP} {\bfseries 05}
  (2023) 211} [\href{https://arxiv.org/abs/2212.07965}{{\ttfamily
  2212.07965}}].

\bibitem{Saketh:2023bul}
M.V.S.~Saketh, Z.~Zhou and M.M.~Ivanov, \emph{{Dynamical tidal response of Kerr
  black holes from scattering amplitudes}},
  \href{https://doi.org/10.1103/PhysRevD.109.064058}{\emph{Phys. Rev. D}
  {\bfseries 109} (2024) 064058}
  [\href{https://arxiv.org/abs/2307.10391}{{\ttfamily 2307.10391}}].

\bibitem{Bautista:2023sdf}
Y.F.~Bautista, G.~Bonelli, C.~Iossa, A.~Tanzini and Z.~Zhou, \emph{{Black Hole
  Perturbation Theory Meets CFT$_2$: Kerr Compton Amplitudes from
  Nekrasov-Shatashvili Functions}},
  \href{https://arxiv.org/abs/2312.05965}{{\ttfamily 2312.05965}}.

\bibitem{Chung:2018kqs}
M.-Z.~Chung, Y.-T.~Huang, J.-W.~Kim and S.~Lee, \emph{{The simplest massive
  S-matrix: from minimal coupling to Black Holes}},
  \href{https://doi.org/10.1007/JHEP04(2019)156}{\emph{JHEP} {\bfseries 04}
  (2019) 156} [\href{https://arxiv.org/abs/1812.08752}{{\ttfamily
  1812.08752}}].

\bibitem{Chung:2019duq}
M.-Z.~Chung, Y.-T.~Huang and J.-W.~Kim, \emph{{Classical potential for general
  spinning bodies}}, \href{https://doi.org/10.1007/JHEP09(2020)074}{\emph{JHEP}
  {\bfseries 09} (2020) 074}
  [\href{https://arxiv.org/abs/1908.08463}{{\ttfamily 1908.08463}}].

\bibitem{Chung:2020rrz}
M.-Z.~Chung, Y.-t.~Huang, J.-W.~Kim and S.~Lee, \emph{{Complete Hamiltonian for
  spinning binary systems at first post-Minkowskian order}},
  \href{https://doi.org/10.1007/JHEP05(2020)105}{\emph{JHEP} {\bfseries 05}
  (2020) 105} [\href{https://arxiv.org/abs/2003.06600}{{\ttfamily
  2003.06600}}].

\bibitem{Chen:2021kxt}
W.-M.~Chen, M.-Z.~Chung, Y.-t.~Huang and J.-W.~Kim, \emph{{The 2PM Hamiltonian
  for binary Kerr to quartic in spin}},
  \href{https://doi.org/10.1007/JHEP08(2022)148}{\emph{JHEP} {\bfseries 08}
  (2022) 148} [\href{https://arxiv.org/abs/2111.13639}{{\ttfamily
  2111.13639}}].

\bibitem{Aoude:2022trd}
R.~Aoude, K.~Haddad and A.~Helset, \emph{{Searching for Kerr in the 2PM
  amplitude}}, \href{https://doi.org/10.1007/JHEP07(2022)072}{\emph{JHEP}
  {\bfseries 07} (2022) 072}
  [\href{https://arxiv.org/abs/2203.06197}{{\ttfamily 2203.06197}}].

\bibitem{Aoude:2022thd}
R.~Aoude, K.~Haddad and A.~Helset, \emph{{Classical Gravitational
  Spinning-Spinless Scattering at O(G2S\ensuremath{\infty})}},
  \href{https://doi.org/10.1103/PhysRevLett.129.141102}{\emph{Phys. Rev. Lett.}
  {\bfseries 129} (2022) 141102}
  [\href{https://arxiv.org/abs/2205.02809}{{\ttfamily 2205.02809}}].

\bibitem{Bern:2022kto}
Z.~Bern, D.~Kosmopoulos, A.~Luna, R.~Roiban and F.~Teng, \emph{{Binary Dynamics
  Through the Fifth Power of Spin at $\mathcal{O}(G^2)$}},
  \href{https://arxiv.org/abs/2203.06202}{{\ttfamily 2203.06202}}.

\bibitem{Bern:2023ity}
Z.~Bern, D.~Kosmopoulos, A.~Luna, R.~Roiban, T.~Scheopner, F.~Teng et~al.,
  \emph{{Quantum Field Theory, Worldline Theory, and Spin Magnitude Change in
  Orbital Evolution}},  \href{https://arxiv.org/abs/2308.14176}{{\ttfamily
  2308.14176}}.

\bibitem{Menezes:2022tcs}
G.~Menezes and M.~Sergola, \emph{{NLO deflections for spinning particles and
  Kerr black holes}},
  \href{https://doi.org/10.1007/JHEP10(2022)105}{\emph{JHEP} {\bfseries 10}
  (2022) 105} [\href{https://arxiv.org/abs/2205.11701}{{\ttfamily
  2205.11701}}].

\bibitem{FebresCordero:2022jts}
F.~Febres~Cordero, M.~Kraus, G.~Lin, M.S.~Ruf and M.~Zeng, \emph{{Conservative
  Binary Dynamics with a Spinning Black Hole at O(G3) from Scattering
  Amplitudes}},
  \href{https://doi.org/10.1103/PhysRevLett.130.021601}{\emph{Phys. Rev. Lett.}
  {\bfseries 130} (2023) 021601}
  [\href{https://arxiv.org/abs/2205.07357}{{\ttfamily 2205.07357}}].

\bibitem{Cangemi:2022bew}
L.~Cangemi, M.~Chiodaroli, H.~Johansson, A.~Ochirov, P.~Pichini and
  E.~Skvortsov, \emph{{Kerr Black Holes Enjoy Massive Higher-Spin Gauge
  Symmetry}},  \href{https://arxiv.org/abs/2212.06120}{{\ttfamily 2212.06120}}.

\bibitem{Haddad:2023ylx}
K.~Haddad, \emph{{Recursion in the classical limit and the neutron-star Compton
  amplitude}}, \href{https://doi.org/10.1007/JHEP05(2023)177}{\emph{JHEP}
  {\bfseries 05} (2023) 177}
  [\href{https://arxiv.org/abs/2303.02624}{{\ttfamily 2303.02624}}].

\bibitem{DeAngelis:2023lvf}
S.~De~Angelis, R.~Gonzo and P.P.~Novichkov, \emph{{Spinning waveforms from KMOC
  at leading order}},  \href{https://arxiv.org/abs/2309.17429}{{\ttfamily
  2309.17429}}.

\bibitem{Chen:2023qzo}
Y.-J.~Chen, T.~Hsieh, Y.-T.~Huang and J.-W.~Kim, \emph{{On-shell approach to
  (spinning) gravitational absorption processes}},
  \href{https://arxiv.org/abs/2312.04513}{{\ttfamily 2312.04513}}.

\bibitem{Cangemi:2023ysz}
L.~Cangemi, M.~Chiodaroli, H.~Johansson, A.~Ochirov, P.~Pichini and
  E.~Skvortsov, \emph{{From higher-spin gauge interactions to Compton
  amplitudes for root-Kerr}},
  \href{https://arxiv.org/abs/2311.14668}{{\ttfamily 2311.14668}}.

\bibitem{Cangemi:2023bpe}
L.~Cangemi, M.~Chiodaroli, H.~Johansson, A.~Ochirov, P.~Pichini and
  E.~Skvortsov, \emph{{Compton Amplitude for Rotating Black Hole from QFT}},
  \href{https://arxiv.org/abs/2312.14913}{{\ttfamily 2312.14913}}.

\bibitem{Vines:2017hyw}
J.~Vines, \emph{{Scattering of two spinning black holes in post-Minkowskian
  gravity, to all orders in spin, and effective-one-body mappings}},
  \href{https://doi.org/10.1088/1361-6382/aaa3a8}{\emph{Class. Quant. Grav.}
  {\bfseries 35} (2018) 084002}
  [\href{https://arxiv.org/abs/1709.06016}{{\ttfamily 1709.06016}}].

\bibitem{Guevara:2018wpp}
A.~Guevara, A.~Ochirov and J.~Vines, \emph{{Scattering of Spinning Black Holes
  from Exponentiated Soft Factors}},
  \href{https://doi.org/10.1007/JHEP09(2019)056}{\emph{JHEP} {\bfseries 09}
  (2019) 056} [\href{https://arxiv.org/abs/1812.06895}{{\ttfamily
  1812.06895}}].

\bibitem{Kosmopoulos:2021zoq}
D.~Kosmopoulos and A.~Luna, \emph{{Quadratic-in-spin Hamiltonian at $
  \mathcal{O} $(G$^{2}$) from scattering amplitudes}},
  \href{https://doi.org/10.1007/JHEP07(2021)037}{\emph{JHEP} {\bfseries 07}
  (2021) 037} [\href{https://arxiv.org/abs/2102.10137}{{\ttfamily
  2102.10137}}].

\bibitem{Scheopner:2023rzp}
T.~Scheopner and J.~Vines, \emph{{Dynamical Implications of the Kerr Multipole
  Moments for Spinning Black Holes}},
  \href{https://arxiv.org/abs/2311.18421}{{\ttfamily 2311.18421}}.

\bibitem{Alessio:2022kwv}
F.~Alessio and P.~Di~Vecchia, \emph{{Radiation reaction for spinning black-hole
  scattering}},
  \href{https://doi.org/10.1016/j.physletb.2022.137258}{\emph{Phys. Lett. B}
  {\bfseries 832} (2022) 137258}
  [\href{https://arxiv.org/abs/2203.13272}{{\ttfamily 2203.13272}}].

\bibitem{Elkhidir:2023dco}
A.~Elkhidir, D.~O'Connell, M.~Sergola and I.A.~Vazquez-Holm, \emph{{Radiation
  and Reaction at One Loop}},
  \href{https://arxiv.org/abs/2303.06211}{{\ttfamily 2303.06211}}.

\bibitem{Bini:2023fiz}
D.~Bini, T.~Damour and A.~Geralico, \emph{{Comparing one-loop gravitational
  bremsstrahlung amplitudes to the multipolar-post-Minkowskian waveform}},
  \href{https://doi.org/10.1103/PhysRevD.108.124052}{\emph{Phys. Rev. D}
  {\bfseries 108} (2023) 124052}
  [\href{https://arxiv.org/abs/2309.14925}{{\ttfamily 2309.14925}}].

\bibitem{Brandhuber:2023hhl}
A.~Brandhuber, G.R.~Brown, G.~Chen, J.~Gowdy and G.~Travaglini, \emph{{Resummed
  spinning waveforms from five-point amplitudes}},
  \href{https://doi.org/10.1007/JHEP02(2024)026}{\emph{JHEP} {\bfseries 02}
  (2024) 026} [\href{https://arxiv.org/abs/2310.04405}{{\ttfamily
  2310.04405}}].

\bibitem{Luna:2023uwd}
A.~Luna, N.~Moynihan, D.~O'Connell and A.~Ross, \emph{{Observables from the
  Spinning Eikonal}},  \href{https://arxiv.org/abs/2312.09960}{{\ttfamily
  2312.09960}}.

\bibitem{Lee:2023nkx}
H.~Lee and S.~Lee, \emph{{Poincar\'e invariance of spinning binary dynamics in
  the post-Minkowskian Hamiltonian approach}},
  \href{https://doi.org/10.1088/1361-6382/ad0992}{\emph{Class. Quant. Grav.}
  {\bfseries 40} (2023) 245004}
  [\href{https://arxiv.org/abs/2305.10739}{{\ttfamily 2305.10739}}].

\bibitem{Buonanno:2024vkx}
A.~Buonanno, G.U.~Jakobsen and G.~Mogull, \emph{{Post-Minkowskian Theory Meets
  the Spinning Effective-One-Body Approach for Two-Body Scattering}},
  \href{https://arxiv.org/abs/2402.12342}{{\ttfamily 2402.12342}}.

\bibitem{Jakobsen:2021lvp}
G.U.~Jakobsen, G.~Mogull, J.~Plefka and J.~Steinhoff, \emph{{Gravitational
  Bremsstrahlung and Hidden Supersymmetry of Spinning Bodies}},
  \href{https://arxiv.org/abs/2106.10256}{{\ttfamily 2106.10256}}.

\bibitem{Jakobsen:2021zvh}
G.U.~Jakobsen, G.~Mogull, J.~Plefka and J.~Steinhoff, \emph{{SUSY in the sky
  with gravitons}}, \href{https://doi.org/10.1007/JHEP01(2022)027}{\emph{JHEP}
  {\bfseries 01} (2022) 027}
  [\href{https://arxiv.org/abs/2109.04465}{{\ttfamily 2109.04465}}].

\bibitem{Jakobsen:2022zsx}
G.U.~Jakobsen and G.~Mogull, \emph{{Linear response, Hamiltonian, and radiative
  spinning two-body dynamics}},
  \href{https://doi.org/10.1103/PhysRevD.107.044033}{\emph{Phys. Rev. D}
  {\bfseries 107} (2023) 044033}
  [\href{https://arxiv.org/abs/2210.06451}{{\ttfamily 2210.06451}}].

\bibitem{Jakobsen:2023ndj}
G.U.~Jakobsen, G.~Mogull, J.~Plefka, B.~Sauer and Y.~Xu, \emph{{Conservative
  scattering of spinning black holes at fourth post-Minkowskian order}},
  \href{https://arxiv.org/abs/2306.01714}{{\ttfamily 2306.01714}}.

\bibitem{Jakobsen:2023hig}
G.U.~Jakobsen, G.~Mogull, J.~Plefka and B.~Sauer, \emph{{Dissipative Scattering
  of Spinning Black Holes at Fourth Post-Minkowskian Order}},
  \href{https://doi.org/10.1103/PhysRevLett.131.241402}{\emph{Phys. Rev. Lett.}
  {\bfseries 131} (2023) 241402}
  [\href{https://arxiv.org/abs/2308.11514}{{\ttfamily 2308.11514}}].

\bibitem{Shi:2021qsb}
C.~Shi and J.~Plefka, \emph{{Classical double copy of worldline quantum field
  theory}}, \href{https://doi.org/10.1103/PhysRevD.105.026007}{\emph{Phys. Rev.
  D} {\bfseries 105} (2022) 026007}
  [\href{https://arxiv.org/abs/2109.10345}{{\ttfamily 2109.10345}}].

\bibitem{Comberiati:2022cpm}
F.~Comberiati and C.~Shi, \emph{{Classical Double Copy of Spinning Worldline
  Quantum Field Theory}},
  \href{https://doi.org/10.1007/JHEP04(2023)008}{\emph{JHEP} {\bfseries 04}
  (2023) 008} [\href{https://arxiv.org/abs/2212.13855}{{\ttfamily
  2212.13855}}].

\bibitem{Gonzo:2023goe}
R.~Gonzo and C.~Shi, \emph{{Boundary to bound dictionary for generic Kerr
  orbits}}, \href{https://doi.org/10.1103/PhysRevD.108.084065}{\emph{Phys. Rev.
  D} {\bfseries 108} (2023) 084065}
  [\href{https://arxiv.org/abs/2304.06066}{{\ttfamily 2304.06066}}].

\bibitem{Gonzo:2024zxo}
R.~Gonzo and C.~Shi, \emph{{Scattering and bound observables for spinning
  particles in Kerr spacetime with generic spin orientations}},
  \href{https://arxiv.org/abs/2405.09687}{{\ttfamily 2405.09687}}.

\bibitem{Aoude:2023dui}
R.~Aoude, K.~Haddad, C.~Heissenberg and A.~Helset, \emph{{Leading-order
  gravitational radiation to all spin orders}},
  \href{https://doi.org/10.1103/PhysRevD.109.036007}{\emph{Phys. Rev. D}
  {\bfseries 109} (2024) 036007}
  [\href{https://arxiv.org/abs/2310.05832}{{\ttfamily 2310.05832}}].

\bibitem{Bohnenblust:2023qmy}
L.~Bohnenblust, H.~Ita, M.~Kraus and J.~Schlenk, \emph{{Gravitational
  Bremsstrahlung in Black-Hole Scattering at $\mathcal{O}(G^3)$: Linear-in-Spin
  Effects}},  \href{https://arxiv.org/abs/2312.14859}{{\ttfamily 2312.14859}}.

\bibitem{Skvortsov:2023jbn}
E.~Skvortsov and M.~Tsulaia, \emph{{Cubic action for spinning black holes from
  massive higher-spin gauge symmetry}},
  \href{https://doi.org/10.1007/JHEP02(2024)202}{\emph{JHEP} {\bfseries 02}
  (2024) 202} [\href{https://arxiv.org/abs/2312.08184}{{\ttfamily
  2312.08184}}].

\bibitem{Kim:2024grz}
J.-H.~Kim, J.-W.~Kim and S.~Lee, \emph{{Massive twistor worldline in
  electromagnetic fields}},  \href{https://arxiv.org/abs/2405.17056}{{\ttfamily
  2405.17056}}.

\bibitem{Brandhuber:2024bnz}
A.~Brandhuber, G.R.~Brown, P.~Pichini, G.~Travaglini and P.V.~Matasan,
  \emph{{Spinning binary dynamics in cubic effective field theories of
  gravity}},  \href{https://arxiv.org/abs/2405.13826}{{\ttfamily 2405.13826}}.

\bibitem{Bern:2024adl}
Z.~Bern, E.~Herrmann, R.~Roiban, M.S.~Ruf, A.V.~Smirnov, V.A.~Smirnov et~al.,
  \emph{{Amplitudes, Supersymmetric Black Hole Scattering at
  $\mathcal{O}(G^5)$, and Loop Integration}},
  \href{https://arxiv.org/abs/2406.01554}{{\ttfamily 2406.01554}}.

\bibitem{Kalin:2020fhe}
G.~K\"alin, Z.~Liu and R.A.~Porto, \emph{{Conservative Dynamics of Binary
  Systems to Third Post-Minkowskian Order from the Effective Field Theory
  Approach}}, \href{https://doi.org/10.1103/PhysRevLett.125.261103}{\emph{Phys.
  Rev. Lett.} {\bfseries 125} (2020) 261103}
  [\href{https://arxiv.org/abs/2007.04977}{{\ttfamily 2007.04977}}].

\bibitem{Kalin:2020lmz}
G.~K\"alin, Z.~Liu and R.A.~Porto, \emph{{Conservative Tidal Effects in Compact
  Binary Systems to Next-to-Leading Post-Minkowskian Order}},
  \href{https://doi.org/10.1103/PhysRevD.102.124025}{\emph{Phys. Rev. D}
  {\bfseries 102} (2020) 124025}
  [\href{https://arxiv.org/abs/2008.06047}{{\ttfamily 2008.06047}}].

\bibitem{Riva:2021vnj}
M.M.~Riva and F.~Vernizzi, \emph{{Radiated momentum in the post-Minkowskian
  worldline approach via reverse unitarity}},
  \href{https://doi.org/10.1007/JHEP11(2021)228}{\emph{JHEP} {\bfseries 11}
  (2021) 228} [\href{https://arxiv.org/abs/2110.10140}{{\ttfamily
  2110.10140}}].

\bibitem{Dlapa:2021vgp}
C.~Dlapa, G.~K\"alin, Z.~Liu and R.A.~Porto, \emph{{Conservative Dynamics of
  Binary Systems at Fourth Post-Minkowskian Order in the Large-Eccentricity
  Expansion}},
  \href{https://doi.org/10.1103/PhysRevLett.128.161104}{\emph{Phys. Rev. Lett.}
  {\bfseries 128} (2022) 161104}
  [\href{https://arxiv.org/abs/2112.11296}{{\ttfamily 2112.11296}}].

\bibitem{Goldberger:2020fot}
W.D.~Goldberger, J.~Li and I.Z.~Rothstein, \emph{{Non-conservative effects on
  spinning black holes from world-line effective field theory}},
  \href{https://doi.org/10.1007/JHEP06(2021)053}{\emph{JHEP} {\bfseries 06}
  (2021) 053} [\href{https://arxiv.org/abs/2012.14869}{{\ttfamily
  2012.14869}}].

\bibitem{Kalin:2020mvi}
G.~K\"{a}lin and R.A.~Porto, \emph{{Post-Minkowskian Effective Field Theory for
  Conservative Binary Dynamics}},
  \href{https://arxiv.org/abs/2006.01184}{{\ttfamily 2006.01184}}.

\bibitem{Bhattacharyya:2024aeq}
A.~Bhattacharyya, D.~Ghosh, S.~Ghosh and S.~Pal, \emph{{Observables from
  classical black hole scattering in Scalar-Tensor theory of gravity from
  worldline quantum field theory}},
  \href{https://doi.org/10.1007/JHEP04(2024)015}{\emph{JHEP} {\bfseries 04}
  (2024) 015} [\href{https://arxiv.org/abs/2401.05492}{{\ttfamily
  2401.05492}}].

\bibitem{Kosmopoulos:2023bwc}
D.~Kosmopoulos and M.P.~Solon, \emph{{Gravitational self force from scattering
  amplitudes in curved space}},
  \href{https://doi.org/10.1007/JHEP03(2024)125}{\emph{JHEP} {\bfseries 03}
  (2024) 125} [\href{https://arxiv.org/abs/2308.15304}{{\ttfamily
  2308.15304}}].

\bibitem{Cheung:2023lnj}
C.~Cheung, J.~Parra-Martinez, I.Z.~Rothstein, N.~Shah and J.~Wilson-Gerow,
  \emph{{Effective Field Theory for Extreme Mass Ratios}},
  \href{https://arxiv.org/abs/2308.14832}{{\ttfamily 2308.14832}}.

\bibitem{Wilson-Gerow:2023syq}
J.~Wilson-Gerow, \emph{{Conservative scattering of Reissner-Nordstr\"om black
  holes at third post-Minkowskian order}},
  \href{https://doi.org/10.1007/JHEP05(2024)265}{\emph{JHEP} {\bfseries 05}
  (2024) 265} [\href{https://arxiv.org/abs/2310.17731}{{\ttfamily
  2310.17731}}].

\bibitem{Jakobsen:2023tvm}
G.U.~Jakobsen, \emph{{Spin and Susceptibility Effects of Electromagnetic
  Self-Force in Effective Field Theory}},
  \href{https://doi.org/10.1103/PhysRevLett.132.151601}{\emph{Phys. Rev. Lett.}
  {\bfseries 132} (2024) 151601}
  [\href{https://arxiv.org/abs/2311.04151}{{\ttfamily 2311.04151}}].

\bibitem{Klemm:2024wtd}
A.~Klemm, C.~Nega, B.~Sauer and J.~Plefka, \emph{{CY in the Sky}},
  \href{https://arxiv.org/abs/2401.07899}{{\ttfamily 2401.07899}}.

\bibitem{Arkani-Hamed:2019ymq}
N.~Arkani-Hamed, Y.-t.~Huang and D.~O'Connell, \emph{{Kerr black holes as
  elementary particles}},
  \href{https://doi.org/10.1007/JHEP01(2020)046}{\emph{JHEP} {\bfseries 01}
  (2020) 046} [\href{https://arxiv.org/abs/1906.10100}{{\ttfamily
  1906.10100}}].

\bibitem{Johansson:2019dnu}
H.~Johansson and A.~Ochirov, \emph{{Double copy for massive quantum particles
  with spin}}, \href{https://doi.org/10.1007/JHEP09(2019)040}{\emph{JHEP}
  {\bfseries 09} (2019) 040}
  [\href{https://arxiv.org/abs/1906.12292}{{\ttfamily 1906.12292}}].

\bibitem{Georgi:1990um}
H.~Georgi, \emph{{An Effective Field Theory for Heavy Quarks at Low-energies}},
  \href{https://doi.org/10.1016/0370-2693(90)91128-X}{\emph{Phys. Lett. B}
  {\bfseries 240} (1990) 447}.

\bibitem{Luke:1992cs}
M.E.~Luke and A.V.~Manohar, \emph{{Reparametrization invariance constraints on
  heavy particle effective field theories}},
  \href{https://doi.org/10.1016/0370-2693(92)91786-9}{\emph{Phys. Lett. B}
  {\bfseries 286} (1992) 348}
  [\href{https://arxiv.org/abs/hep-ph/9205228}{{\ttfamily hep-ph/9205228}}].

\bibitem{Neubert:1993mb}
M.~Neubert, \emph{{Heavy quark symmetry}},
  \href{https://doi.org/10.1016/0370-1573(94)90091-4}{\emph{Phys. Rept.}
  {\bfseries 245} (1994) 259}
  [\href{https://arxiv.org/abs/hep-ph/9306320}{{\ttfamily hep-ph/9306320}}].

\bibitem{Manohar:2000dt}
A.V.~Manohar and M.B.~Wise, \emph{{Heavy quark physics}}, vol.~10, Cambridge
  University Press (2000).

\bibitem{Damgaard:2019lfh}
P.H.~Damgaard, K.~Haddad and A.~Helset, \emph{{Heavy Black Hole Effective
  Theory}}, \href{https://doi.org/10.1007/JHEP11(2019)070}{\emph{JHEP}
  {\bfseries 11} (2019) 070}
  [\href{https://arxiv.org/abs/1908.10308}{{\ttfamily 1908.10308}}].

\bibitem{Conde:2016izb}
E.~Conde, E.~Joung and K.~Mkrtchyan, \emph{{Spinor-Helicity Three-Point
  Amplitudes from Local Cubic Interactions}},
  \href{https://doi.org/10.1007/JHEP08(2016)040}{\emph{JHEP} {\bfseries 08}
  (2016) 040} [\href{https://arxiv.org/abs/1605.07402}{{\ttfamily
  1605.07402}}].

\bibitem{Conde:2016vxs}
E.~Conde and A.~Marzolla, \emph{{Lorentz Constraints on Massive Three-Point
  Amplitudes}}, \href{https://doi.org/10.1007/JHEP09(2016)041}{\emph{JHEP}
  {\bfseries 09} (2016) 041}
  [\href{https://arxiv.org/abs/1601.08113}{{\ttfamily 1601.08113}}].

\bibitem{Brandhuber:2021bsf}
A.~Brandhuber, G.~Chen, H.~Johansson, G.~Travaglini and C.~Wen,
  \emph{{Kinematic Hopf Algebra for Bern-Carrasco-Johansson Numerators in
  Heavy-Mass Effective Field Theory and Yang-Mills Theory}},
  \href{https://doi.org/10.1103/PhysRevLett.128.121601}{\emph{Phys. Rev. Lett.}
  {\bfseries 128} (2022) 121601}
  [\href{https://arxiv.org/abs/2111.15649}{{\ttfamily 2111.15649}}].

\bibitem{Brandhuber:2022enp}
A.~Brandhuber, G.R.~Brown, G.~Chen, J.~Gowdy, G.~Travaglini and C.~Wen,
  \emph{{Amplitudes, Hopf algebras and the colour-kinematics duality}},
  \href{https://doi.org/10.1007/JHEP12(2022)101}{\emph{JHEP} {\bfseries 12}
  (2022) 101} [\href{https://arxiv.org/abs/2208.05886}{{\ttfamily
  2208.05886}}].

\bibitem{Chen:2022nei}
G.~Chen, G.~Lin and C.~Wen, \emph{{Kinematic Hopf algebra for amplitudes and
  form factors}},
  \href{https://doi.org/10.1103/PhysRevD.107.L081701}{\emph{Phys. Rev. D}
  {\bfseries 107} (2023) L081701}
  [\href{https://arxiv.org/abs/2208.05519}{{\ttfamily 2208.05519}}].

\bibitem{Chen:2023ekh}
G.~Chen, L.~Rodina and C.~Wen, \emph{{Kinematic Hopf algebra for amplitudes
  from higher-derivative operators}},
  \href{https://doi.org/10.1007/JHEP02(2024)096}{\emph{JHEP} {\bfseries 02}
  (2024) 096} [\href{https://arxiv.org/abs/2310.11943}{{\ttfamily
  2310.11943}}].

\bibitem{Chen:2024gkj}
G.~Chen, L.~Rodina and C.~Wen, \emph{{Kinematic Hopf algebra and BCJ numerators
  at finite $\alpha'$}},  \href{https://arxiv.org/abs/2403.04614}{{\ttfamily
  2403.04614}}.

\bibitem{Bjerrum-Bohr:2024fbt}
N.E.J.~Bjerrum-Bohr, G.~Chen, Y.~Miao and M.~Skowronek, \emph{{Color-Kinematic
  Numerators for Fermion Compton Amplitudes}},
  \href{https://arxiv.org/abs/2404.15265}{{\ttfamily 2404.15265}}.

\bibitem{Chen:2019ywi}
G.~Chen, H.~Johansson, F.~Teng and T.~Wang, \emph{{On the kinematic algebra for
  BCJ numerators beyond the MHV sector}},
  \href{https://doi.org/10.1007/JHEP11(2019)055}{\emph{JHEP} {\bfseries 11}
  (2019) 055} [\href{https://arxiv.org/abs/1906.10683}{{\ttfamily
  1906.10683}}].

\bibitem{Chen:2021chy}
G.~Chen, H.~Johansson, F.~Teng and T.~Wang, \emph{{Next-to-MHV Yang-Mills
  kinematic algebra}},
  \href{https://doi.org/10.1007/JHEP10(2021)042}{\emph{JHEP} {\bfseries 10}
  (2021) 042} [\href{https://arxiv.org/abs/2104.12726}{{\ttfamily
  2104.12726}}].

\bibitem{Bautista:2023szu}
Y.F.~Bautista, \emph{{Dynamics for Super-Extremal Kerr Binary Systems at ${\cal
  O}(G^2)$}},  \href{https://arxiv.org/abs/2304.04287}{{\ttfamily 2304.04287}}.

\bibitem{Dolan:2008kf}
S.R.~Dolan, \emph{{Scattering and Absorption of Gravitational Plane Waves by
  Rotating Black Holes}},
  \href{https://doi.org/10.1088/0264-9381/25/23/235002}{\emph{Class. Quant.
  Grav.} {\bfseries 25} (2008) 235002}
  [\href{https://arxiv.org/abs/0801.3805}{{\ttfamily 0801.3805}}].

\bibitem{Kim:2023drc}
J.-W.~Kim and J.~Steinhoff, \emph{{Spin supplementary condition in quantum
  field theory: covariant SSC and physical state projection}},
  \href{https://doi.org/10.1007/JHEP07(2023)042}{\emph{JHEP} {\bfseries 07}
  (2023) 042} [\href{https://arxiv.org/abs/2302.01944}{{\ttfamily
  2302.01944}}].

\bibitem{Jakobsen:2022psy}
G.U.~Jakobsen, G.~Mogull, J.~Plefka and B.~Sauer, \emph{{All things retarded:
  radiation-reaction in worldline quantum field theory}},
  \href{https://doi.org/10.1007/JHEP10(2022)128}{\emph{JHEP} {\bfseries 10}
  (2022) 128} [\href{https://arxiv.org/abs/2207.00569}{{\ttfamily
  2207.00569}}].

\bibitem{Witzany:2019nml}
V.~Witzany, \emph{{Hamilton-Jacobi equation for spinning particles near black
  holes}}, \href{https://doi.org/10.1103/PhysRevD.100.104030}{\emph{Phys. Rev.
  D} {\bfseries 100} (2019) 104030}
  [\href{https://arxiv.org/abs/1903.03651}{{\ttfamily 1903.03651}}].

\bibitem{Bjerrum-Bohr:2021vuf}
N.E.J.~Bjerrum-Bohr, P.H.~Damgaard, L.~Plant\'e and P.~Vanhove,
  \emph{{Classical gravity from loop amplitudes}},
  \href{https://doi.org/10.1103/PhysRevD.104.026009}{\emph{Phys. Rev. D}
  {\bfseries 104} (2021) 026009}
  [\href{https://arxiv.org/abs/2104.04510}{{\ttfamily 2104.04510}}].

\bibitem{Bjerrum-Bohr:2021din}
N.E.J.~Bjerrum-Bohr, P.H.~Damgaard, L.~Plant\'e and P.~Vanhove, \emph{{The
  Amplitude for Classical Gravitational Scattering at Third Post-Minkowskian
  Order}},  \href{https://arxiv.org/abs/2105.05218}{{\ttfamily 2105.05218}}.

\bibitem{Bern:1994zx}
Z.~Bern, L.J.~Dixon, D.C.~Dunbar and D.A.~Kosower, \emph{{One loop $n$-point
  gauge theory amplitudes, unitarity and collinear limits}},
  \href{https://doi.org/10.1016/0550-3213(94)90179-1}{\emph{Nucl. Phys.}
  {\bfseries B425} (1994) 217}
  [\href{https://arxiv.org/abs/hep-ph/9403226}{{\ttfamily hep-ph/9403226}}].

\bibitem{Bern:1994cg}
Z.~Bern, L.J.~Dixon, D.C.~Dunbar and D.A.~Kosower, \emph{{Fusing gauge theory
  tree amplitudes into loop amplitudes}},
  \href{https://doi.org/10.1016/0550-3213(94)00488-Z}{\emph{Nucl. Phys.}
  {\bfseries B435} (1995) 59}
  [\href{https://arxiv.org/abs/hep-ph/9409265}{{\ttfamily hep-ph/9409265}}].

\bibitem{Bjerrum-Bohr:2013bxa}
N.E.J.~Bjerrum-Bohr, J.F.~Donoghue and P.~Vanhove, \emph{{On-shell Techniques
  and Universal Results in Quantum Gravity}},
  \href{https://doi.org/10.1007/JHEP02(2014)111}{\emph{JHEP} {\bfseries 02}
  (2014) 111} [\href{https://arxiv.org/abs/1309.0804}{{\ttfamily 1309.0804}}].

\bibitem{Cachazo:2017jef}
F.~Cachazo and A.~Guevara, \emph{{Leading Singularities and Classical
  Gravitational Scattering}},
  \href{https://doi.org/10.1007/JHEP02(2020)181}{\emph{JHEP} {\bfseries 02}
  (2020) 181} [\href{https://arxiv.org/abs/1705.10262}{{\ttfamily
  1705.10262}}].

\bibitem{KiHA}
G.~Chen, \emph{{Kinematic Hopf Algebra}},
  \href{https://arxiv.org/abs/https://github.com/AmplitudeGravity/kinematicHopfAlgebra}{{\ttfamily
  https://github.com/AmplitudeGravity/kinematicHopfAlgebra}}.

\bibitem{Lee:2012cn}
R.N.~Lee, \emph{{Presenting LiteRed: a tool for the Loop InTEgrals REDuction}},
   \href{https://arxiv.org/abs/1212.2685}{{\ttfamily 1212.2685}}.

\bibitem{Lee:2013mka}
R.N.~Lee, \emph{{LiteRed 1.4: a powerful tool for reduction of multiloop
  integrals}}, \href{https://doi.org/10.1088/1742-6596/523/1/012059}{\emph{J.
  Phys. Conf. Ser.} {\bfseries 523} (2014) 012059}
  [\href{https://arxiv.org/abs/1310.1145}{{\ttfamily 1310.1145}}].

\bibitem{Bohnenblust:2024hkw}
L.~Bohnenblust, L.~Cangemi, H.~Johansson and P.~Pichini, \emph{{Binary Kerr
  black-hole scattering at 2PM from quantum higher-spin Compton}},
  \href{https://arxiv.org/abs/2410.23271}{{\ttfamily 2410.23271}}.

\end{thebibliography}\endgroup

\end{document}